\documentclass{ametsocV6}

\nolinenumbers 

\title{The modulation of vortex growth by periodic convective activity}

\authors{Hao Fu\aff{a}\aff{b}\aff{c}\correspondingauthor{Hao Fu, haofu@nju.edu.cn}
}

\affiliation{
\aff{a}{School of the Atmospheric Sciences, Nanjing University, Jiangsu, China}\\
\aff{b}{Department of the Geophysical Sciences, University of Chicago, Illinois}\\
\aff{c}{Department of Earth System Science, Stanford University, California}\\
}

\abstract{An important process in tropical cyclone formation is the development of a deep, warm core, which corresponds to the growth of a barotropic cyclone. Persistent convective activity is known to be crucial for the growth of barotropic vorticity. However, it remains unclear whether the fluctuating component of convective activity, such as that caused by the diurnal cycle and inertial-gravity waves, also accelerates the vortex development. To investigate this problem, numerical simulations are performed in an axisymmetric model with the Boussinesq approximation on the f-plane. Convection is parameterized with a bulk-plume mass-flux scheme. To represent a mesoscale convective system modulated by the diurnal cycle, periodic convective mass flux is imposed in a local region. The convection induces periodic diabatic heating and convective momentum transfer in the vertical direction (CMT). The CMT is an irreversible effect that breaks the quadrature phase relation between vertical velocity and vertical vorticity, producing a residual barotropic vorticity in each cycle. The barotropic vorticity consists of a barotropic cyclonic core and an anticyclonic shell. The cyclonic core is produced by the vertical advection and stretching of vertical vorticity. The anticyclonic shell is produced by the radial advection and tilting of radial vorticity. The analytical solution reproduces the formation and growth of the core-shell vorticity structure. This research reveals a potential acceleration effect of periodic convective activity on tropical cyclone genesis. }

\begin{document}

\maketitle

%
%
%
%
%


\section{Introduction} \label{sec_mech:introduction}

The formation of a tropical cyclone (TC) depends crucially on the growth of a deep warm core, which corresponds to a deep cyclonic circulation \citep{palmen1948formation,gray1967mutual,kerns2015subsidence,emanuel2018review}. A crucial component of the deep cyclonic circulation is its vertically uniform aspect, known as the barotropic circulation. The barotropic vorticity accelerates TC formation in many ways. First, it enhances inertial stability, which traps more released latent heat in the inner core of the vortex \citep{hack1986nonlinear}. Second, it increases the surface wind speed, which enhances the wind-induced surface heat flux (WISHE) that warms and humidifies the vortex core \citep{Ooyama_1969,emanuel1986air}.

The barotropic circulation can be produced in at least two ways. The first way is the alignment of a low-level and an upper-level cyclone in a vertically shear flow \citep{dunkerton2009tropical}. The second way is persistent diabatic heating at the same horizontal location, which induces persistent vertical motion \citep{schubert1987PV,majda2008vertically,murthy2019depression,fu2024RRB,fu2025role}. While the stretching of planetary vorticity by an updraft only produces a baroclinic vorticity dipole (a low-level cyclone and an upper-level anticyclone), the vertical advection and stretching of vertical vorticity, which are nonlinear effects, generate a barotropic cyclone at the core of the vertical motion \citep{fu2025role}. Note that the barotropic cyclonic core can also be generated by a downdraft, so long as the vertical motion is persistent \citep{fu2025role}. This paper seeks the third possible way: Does the \textit{fluctuating} component of convective activity also contribute to the vortex growth?

Convective activity in TC genesis is far from smooth, either in the form of random or regular fluctuations \citep{ooyama1982conceptual}. \citet{zehr1992tropical} showed that two externally-driven surges of convection, with an interval of a few days, could precondition TC genesis. The time interval was vaguely argued to be necessary for the vortex structure to adjust, allowing the next convective surge to work more efficiently. Such an external driver may be viewed as a random factor. For regular factors, the diurnal cycle is possibly the most significant one. It influences the convective activity via the periodic shortwave radiative heating of the atmosphere and the periodic modulation of the sea surface temperature \citep{gray1977diurnal,navarro2017balanced,dunion2019tropical,haerter2020diurnal}. Near the circulation center of Hurricane Karl (2010), a rainfall rate undulation of 5 mm hour$^{-1}$ has been reported by \citet{davis2012mesoscale}. This indicates that the influence of the diurnal cycle on TC is amplified by convection, with the variability in latent heating far exceeding the radiative heating. Still for Hurricane Karl (2010), \citet{bell2019mesoscale} reported alternation between low-level and mid-level circulation anomalies, which are associated with deep convective and stratiform clouds and are phase-locked to the diurnal cycle. \citet{wing2022acceleration} performed a carefully designed experiment without diurnal cycle that constrains the total solar insolution to be unchanged. The TC genesis time is found to be slightly advanced. Another fluctuating factor is the inertial-gravity wave, which spans the frequency range between the buoyancy frequency ($\sim$10 minutes scale) and the Coriolis frequency ($\sim$1.5 days in a typical TC genesis latitude of 20$^\circ$N) \citep{o2017accessible}. The wave is particularly relevant for TC genesis in idealized numerical simulations, where a 1000 km-scale doubly periodic domain is typically used. Without a lateral sponge layer, waves travel in and out, producing standing wave-like perturbations \citep{bretherton2005energy,nolan2007trigger,yang2020interactive,fu2023small}. \citet{fu2023small} showed evidence that the standing inertial-gravity wave can excite a persistent vortex, with the vortex size comparable to the wave's half-wavelength.

A complete answer to the posed question involves the two-way coupling between convection and the vortex circulation, modulated by the external fluctuating factor (e.g, the diurnal cycle). To simplify the problem, this paper focuses on a mesoscale convective system, specifically on the response of the circulation to the prescribed convective activity in a local region. Previous idealized studies have examined the circulation response to a local diabatic heating source \citep{gill1980pattern,liu2004vertical_mode,robinson2008resonant,lane2008vortical,liu2019noninstantaneous,Fu2020random,yang2024internal}. However, the role played by the transport of horizontal momentum by convection, referred to as the convective momentum transfer (CMT), has received less attention. The CMT is analogous to an internal frictional effect in the atmosphere \citep{gray1967mutual,schneider1976friction,lee1984bulk,romps2014rayleigh}, which renders the system irreversible. Such an asymmetry in time inspires us to consider whether some residue occurs after a cycle. To fill the gap, we conduct a simple trial by coupling a bulk-plume convective parameterization scheme to an axisymmetric model. The periodic factor is added to the convective mass flux, which modulates \textit{both} diabatic heating and CMT. The simulations exhibit an intriguing result: a barotropic cyclone gradually builds up near the center of the convective region, accompanied by a barotropic anticyclonic shell. An analytical theory is derived to explain the production mechanism of barotropic vorticity, which turns out to be associated with a phase shift between the vertical velocity and vertical vorticity induced by the CMT. 


The paper is organized as follows. Section \ref{sec:axisymmetric} introduces the axisymmetric model framework, including the dynamics part and the convective parameterization scheme. Section \ref{sec:simulation} reports the simulation result. Section \ref{sec:theory} introduces the analytical theory and the comparison with the simulations. Section \ref{sec:summary} concludes the paper.








\section{The axisymmetric model} \label{sec:axisymmetric}

The aim is to investigate how the circulation responds to periodic convective activity, rather than how the convective activity responds to the circulation. Thus, we treat convection in a relatively crude yet physically explicit way by employing a bulk-plume convective parameterization scheme and utilizing an axisymmetric geometry. Only very early axisymmetric models, which have a few vertical layers, used parameterized convection \cite[e.g.,][]{Ooyama_1969}. Since the 1980s, the advance of computational power has enabled most axisymmetric simulations to resolve clouds directly \cite[e.g.,][]{Rotunno_and_Emanuel_1987,bryan2009maximum,wang2021dynamical}. However, clouds in axisymmetric models appear as a ring \citep{peng2019evolution}, which is unrealistic for a mesoscale convective system studied in this paper. Considering the sensitivity of CMT to the geometry of a convective system \citep{lemone1984momentum}, we re-adopted parameterized convection in the axisymmetric model, aware that it remains far less realistic than state-of-the-art 3D cloud-resolving models \citep{montgomery2020contribution}. An additional consideration is that most atmospheric general circulation models (GCMs) still use parameterized convection \cite[e.g.,][]{zhao2012some,tan2018extended}, so the result in this paper can be used to understand TC genesis in GCMs as well.

\subsection{The dynamics part} \label{subsec:dynamics}

The author codes up a model that solves the axisymmetric Boussinesq equation in cylindrical coordinates. It largely follows the method of \citet{lane2008vortical} but includes the tangential flow and some modifications in the numerical discretization. The radial coordinate is $r$, and the vertical coordinate is $z$. The domain radius is $R$, and the domain height is $H$. All boundaries are stress-free and thermally insulating. 


The velocity vector is $(u,v,w)$, and the vorticity vector is $(\xi,\eta,\zeta)$:
\begin{equation}
    \xi = - \frac{\partial v}{\partial z}, \quad
    \eta = \frac{\partial u}{\partial z} - \frac{\partial w}{\partial r},
    \quad 
    \zeta = \frac{\partial v}{\partial r} + \frac{v}{r}.
\end{equation}
The buoyancy, which has been subtracted a linear background profile, is referred to as $b$. The model employs the vorticity-streamfunction formulation, which implicitly accounts for the pressure effect \citep{chien1979vorticity}. We need to solve the $\eta$, $\zeta$, and $b$ equations:
\begin{equation}  \label{eq:eta}
    \frac{\partial \eta}{\partial t} + \frac{\partial u \eta}{\partial r}
    + \frac{\partial w \eta}{\partial z}
    = \frac{\partial \xi v}{\partial r} + \frac{\partial (\zeta + f) v}{\partial z} - \frac{\partial b}{\partial r} + D_{\eta} - s \eta + C_{\eta},
\end{equation}
\begin{equation}  \label{eq:zeta}
    \frac{\partial \zeta}{\partial t}
    + \frac{1}{r}\frac{\partial}{\partial r} \left[ r u (\zeta+f) \right] = \frac{1}{r}\frac{\partial}{\partial r} \left( r \xi w \right) + D_{\zeta} - s \zeta + C_{\zeta}, 
\end{equation}
\begin{equation}  \label{eq:b}
    \frac{\partial b}{\partial t} + 
    \frac{1}{r}\frac{\partial}{\partial r} \left( r u b \right) + \frac{\partial w b}{\partial z} + N^2 w
    = D_b - s b + Q,
\end{equation}
where $D_{\eta}$, $D_{\zeta}$, and $D_b$ are eddy viscosity and diffusivity terms associated with turbulent mixing, $C_{\eta}$ and $C_{\zeta}$ are convective momentum transfer terms, $s$ is the sponge-layer damping rate, $Q$ is the horizontal anomaly of diabatic buoyancy source, and $N$ is the buoyancy frequency. Because the boundaries are stress-free and thermally insulating, the boundary condition of $\eta$, $\zeta$, and $b$ is:
\begin{equation}
    \eta|_{r=0} = \eta|_{r=R} = \eta|_{z=0} = \eta|_{z=H} = 0.
\end{equation}
\begin{equation}
    \frac{\partial \zeta}{\partial r}|_{r=0} = \zeta|_{r=R}= \frac{\partial \zeta}{\partial z}|_{z=0} = \frac{\partial \zeta}{\partial z}|_{z=H} = 0, 
\end{equation}
\begin{equation}
    \frac{\partial b}{\partial r}|_{r=0} = \frac{\partial b}{\partial r}|_{r=R}= \frac{\partial b}{\partial z}|_{z=0} = \frac{\partial b}{\partial z}|_{z=H}=0. 
\end{equation}
Here, the boundary condition at the axis results from axisymmetry.

The turbulent eddy viscosity and diffusion terms use the Laplacian diffusion:
\begin{equation}
    D_{\eta} = K_h \left[ \frac{1}{r} \frac{\partial}{\partial r} \left( r \frac{\partial \eta}{\partial r} \right) - \frac{\eta}{r^2} \right]
    + K_z \frac{\partial^2 \eta}{\partial z^2},
\end{equation}
\begin{equation}
    D_{\zeta} = K_h \frac{1}{r} \frac{\partial}{\partial r} \left( r \frac{\partial \zeta}{\partial r} \right)
    + K_z \frac{\partial^2 \zeta}{\partial z^2},
\end{equation}
\begin{equation}
    D_b = K_h \frac{1}{r} \frac{\partial}{\partial r} \left( r \frac{\partial b}{\partial r} \right)
    + K_z \frac{\partial^2 b}{\partial z^2}.
\end{equation}
Here, $K_h$ is the horizontal eddy diffusivity, and $K_z$ is the vertical eddy diffusivity. They are prescribed constant quantities. 

To avoid the lateral reflection of gravity waves, we follow \citet{lane2008vortical} to impose a lateral sponge layer. The sponge damping rate $s$ is a function of radius:
\begin{equation}
    s = \frac{1}{\tau_s} \exp \left[ -\frac{(r-R)^2}{L_s^2} \right],
\end{equation}
where $\tau_s$ is the damping timescale, and $L_s$ is the characteristic width of the damping zone.

To invert velocity from vorticity, we define the streamfunctions in the horizontal plane $\chi$ (Lagrangian streamfunction) and in the radial-vertical plane $\psi$ (Stokes streamfunction). We first invert $\chi$ and $\psi$ from $\zeta$ and $\eta$:
\begin{equation}  \label{eq:chi}
    \frac{\partial^2 \chi}{\partial r^2} + \frac{1}{r}\frac{\partial \chi}{\partial r} = \zeta, \quad v|_{r=0} = \frac{\partial \chi}{\partial r}|_{r=0} = 0, \quad \chi|_{r=R} = 0, 
\end{equation}
\begin{equation}  \label{eq:psi}
    \frac{\partial^2 \psi}{\partial z^2} + \frac{\partial^2 \psi}{\partial r^2} - \frac{1}{r}\frac{\partial \psi}{\partial r} = - r \eta, 
    \quad 
    \psi|_{r=0} = \psi|_{r=R} =  \psi|_{z=0} = \psi|_{z=H} = 0,
\end{equation}
and then calculate velocity from $\chi$ and $\psi$:
\begin{equation}
    u = - \frac{1}{r}\frac{\partial \psi}{\partial z},
    \quad
    v = \frac{\partial \chi}{\partial r},   \quad
    w = \frac{1}{r}\frac{\partial \psi}{\partial r}.    
\end{equation}
Due to the divergence-free constraint of the vorticity vector, the prognostic equation of radial vorticity $\xi$ is not needed. It is directly calculated from $\xi = - \partial v/ \partial z$. The detailed numerical discretization method is introduced in Appendix A.

\subsection{The bulk-plume convective parameterization scheme} \label{subsec:parameterization}


To parameterize diabatic heating $Q$ and the convective momentum transfer terms $C_\eta$ and $C_\zeta$, we introduce the bulk-plume convective model. The plume denotes a jet of fluid emitted from a local buoyant source, which entrains fluids from the environment \citep{morton1956turbulent,turner1986plume,Yano_2014}. \citet{1947Entrainment} devised the plume model to study cumulus cloud entrainment. \citet{arakawa1974interaction} turned the plume model into a convective parameterization scheme. The scheme divides a grid box into a convective plume part and the environmental part. The plume consists of an ensemble of clouds with different widths, different cloud top heights, and even different lifecycle stages \citep{arakawa1974interaction,mapes2000toy,yano2012interactions,fu2024synchronization}. Among them, the simplest one is the bulk-plume model, which neglects the heterogeneity within the cloud ensemble \citep{zhang1995sensitivity,gregory2001entrainment,de2013entrainment} and has been used in developing analytical models \citep{singh2013entrain,yano2012ODEs,romps2014analytical}. Given its analytically tractable feature, this research adopts the bulk-plume model. 

Two assumptions are used in the derivations below. First, the cloud momentum is assumed to adjust much more rapidly than the environmental momentum, following \citet{romps2012equivalence}. Therefore, the tendency term in the cloud momentum equation can be dropped. Second, the updraft fractional area is assumed to be much smaller than unity \citep{arakawa1974interaction,robe1996moist}. A detailed derivation of the bulk-plume model from the mass and momentum conservation laws is reviewed in Appendix B. 

The averaged convective mass flux per grid box, $M$ (unit: kg m$^{-2}$ s$^{-1}$), obeys: 
\begin{equation}  \label{eq:M_epsilon}
    \frac{1}{M}\frac{\partial M}{\partial z} = \epsilon - \delta,
\end{equation}
where $\epsilon$ is the fractional entrainment rate (unit: m$^{-1}$), and $\delta$ is the fractional detrainment rate (unit: m$^{-1}$). There are two types of $\epsilon$ and $\delta$:
\begin{equation}  \label{eq:epson_sum}
    \epsilon \equiv \epsilon_{dyn} + \epsilon_{edd},
    \quad
    \delta \equiv \delta_{dyn} + \delta_{edd}.
\end{equation}
Here, the subscript $dyn$ denotes the dynamical fractional entrainment and detrainment rate, which denote the transport associated with the cloud-scale convergent and divergent flow \citep{houghton1951theory,de2013entrainment}. The subscript $edd$ denotes the transport associated with smaller-scale eddies. For a typical deep convective updraft, there is dynamical entrainment at the lower level and dynamical detrainment at the upper level; meanwhile, the eddy entrainment and detrainment rates are comparable to each other throughout the depth \citep{romps2010direct}.

\begin{figure}[h]
\centerline{\centering
\includegraphics[width=1\linewidth]{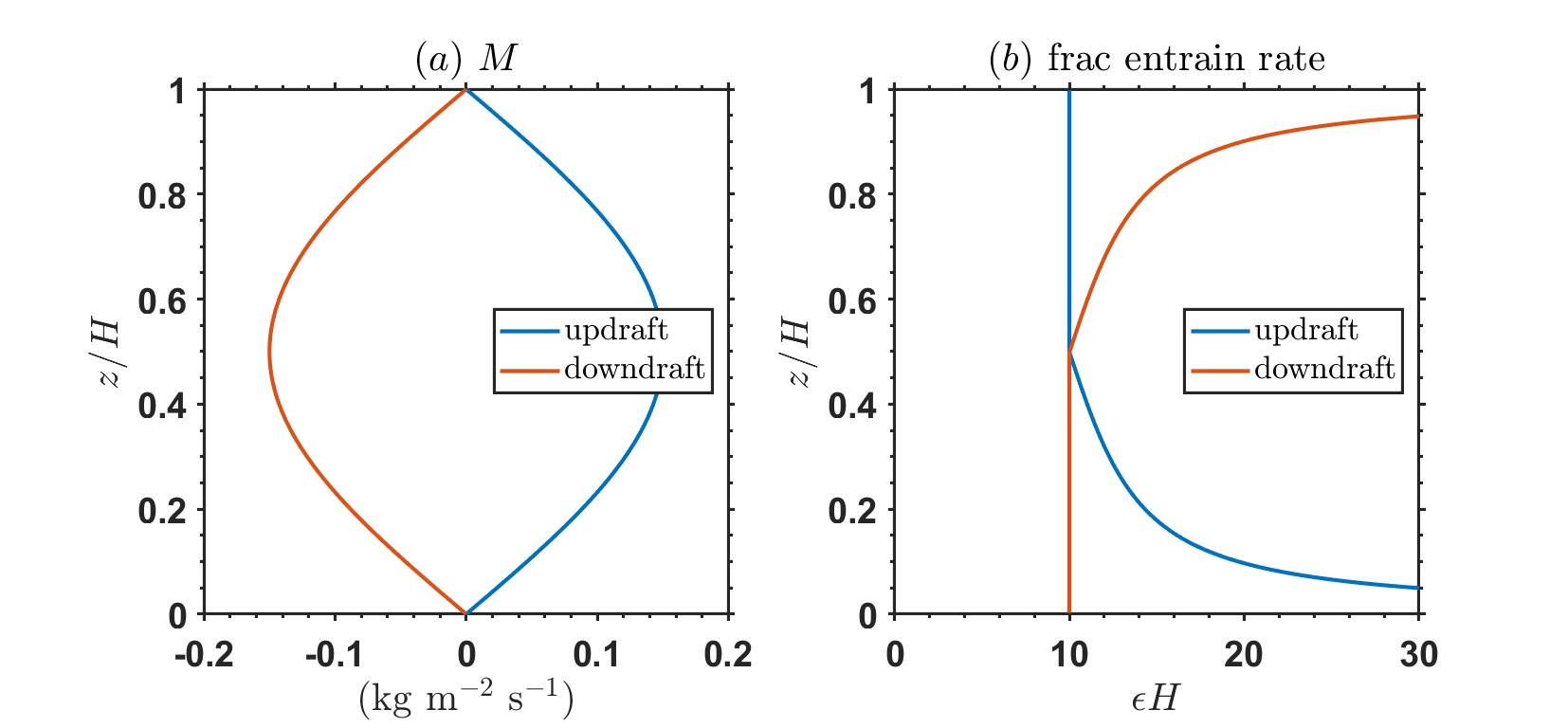}}
  \caption{(a) The vertical profile of convective mass flux $M$ at the domain center, at the peak of the updraft phase (blue line) and at the peak of the downdraft phase (red line). (b) The fractional entrainment rate $\epsilon$ at the peak of the updraft phase (blue line) and at the peak of the downdraft phase (red line). The profiles in (a) only work for the domain center, and the profiles in (b) work uniformly in the domain. }\label{fig:entrain_show}
\end{figure}

In the model, we prescribe $M$ to have the first-baroclinic mode structure in the vertical direction (Fig. \ref{fig:entrain_show}a) and have a cosine track in time:
\begin{equation} \label{eq:M_structure}
    M = M_* \sin \left( \frac{\pi z}{H} \right) \exp \left( - \frac{r^2}{L_M^2} \right) \cos(\Omega t),
\end{equation}
where $M_*$ is the amplitude of the convective mass flux fluctuation, $L_M$ is the horizontal length scale of the convective region, and $\Omega \equiv 2\pi/T$ is the angular frequency of the convective period $T$. The $M$ is fundamentally determined by $\epsilon$ and $\delta$. Based on the diagnostic result of \citet{romps2010direct}, $\epsilon_{edd}$ and $\delta_{edd}$ can be prescribed as a constant:
\begin{equation}  \label{eq:epson_delta_edd}
    \epsilon_{edd} = \delta_{edd} = 1\,\,\mathrm{km^{-1}}.
\end{equation}
Their influence on $M$ cancels out, so $M$ is determined by the dynamical entrainment and detrainment.

The oscillation of $M$ in time represents the alternation between convective updraft and downdraft phases. In the updraft phase, the dynamical entrainment concentrates at the lower level, and the dynamical detrainment concentrates at the upper level. The $\epsilon_{dyn}$ and $\delta_{dyn}$ expressions that comply with Eq. (\ref{eq:M_structure}) are:
\begin{equation}  \label{eq:dyn_updraft}
\mathrm{updraft:}\quad 
    \epsilon_{dyn} = \begin{cases}
        \frac{\pi}{H} \cot \left( \frac{\pi z}{H} \right), \quad z \le \frac{H}{2}, \\
        0, \quad z \ge \frac{H}{2},
    \end{cases}
    \quad 
    \delta_{dyn} = \begin{cases}
        0, \quad z \le \frac{H}{2}, \\
        - \frac{\pi}{H} \cot \left( \frac{\pi z}{H} \right), \quad z \ge \frac{H}{2}.
    \end{cases}
\end{equation}
Note that they only depend on the vertical shape of $M$, rather than the magnitude $M_*$ and the horizontal length scale of the convective region $L_M$. In the downdraft phase, the dynamical entrainment concentrates at the upper level, and the dynamical detrainment concentrates at the lower level:
\begin{equation}  \label{eq:dyn_downdraft}
\mathrm{downdraft:}\quad 
    \epsilon_{dyn} = \begin{cases}
        0, \quad z \le \frac{H}{2}, \\
        - \frac{\pi}{H} \cot \left( \frac{\pi z}{H} \right), \quad z \ge \frac{H}{2},
    \end{cases}
    \quad 
    \delta_{dyn} = \begin{cases}
        \frac{\pi}{H} \cot \left( \frac{\pi z}{H} \right), \quad z \le \frac{H}{2}, \\
        0, \quad z \ge \frac{H}{2}.
    \end{cases}
\end{equation}
See Fig. \ref{fig:entrain_show}b for an illustration of the profile of $\epsilon$ at the updraft and downdraft phases. Note that the updraft and downdraft are approximated as symmetric processes, which facilitates the analytical procedure. However, in the real atmosphere, the shape of vertical velocity profiles in the convective updraft and downdraft differs: \citet{boing2014deceiving} showed that the initiation height of the downdraft is far below the cloud top. Another simplification is the neglect of stratiform heating, a prevalent feature in mesoscale convective systems that exhibits a second-baroclinic-mode heating structure in the vertical direction \citep{houze2004mesoscale}. 


In the bulk-plume model, the diabatic heating is indirectly represented as the local adiabatic warming of the descending flow within the grid box:
\begin{equation}  \label{eq:Q_M}
    Q = \frac{M}{\rho} N^2,
\end{equation}
where the air density $\rho$ is assumed constant. The warming makes the local grid box warmer than the environment. Such a local warming will spread out in the domain via gravity waves, as represented in the axisymmetric model dynamics in section \ref{sec:axisymmetric}\ref{subsec:dynamics}.

Next, we derive CMT in the bulk-plume model. Let the plume-region radial and tangential velocity be $u_c$ and $v_c$. Their evolution depends on the velocity difference between the plume region and the environmental region ($u-u_c$ and $v-v_c$). For an updraft, the in-cloud wind speed equals the environmental wind speed near the surface. As the updraft develops, the cloud entrains air, and the in-cloud wind gets increasingly influenced by the environment. The quasi-steady state cloud momentum equation is:
\begin{equation}  \label{eq:uc_updraft}
\begin{split}
\mathrm{updraft:} \quad
    &\frac{\partial u_c}{\partial z} = \epsilon ( u - u_c ), \quad u_c|_{z=0} = u, \\
    &\frac{\partial v_c}{\partial z} = \epsilon ( v - v_c ), \quad v_c|_{z=0} = v.
\end{split}    
\end{equation}
Similarly, in the downdraft phase, the in-cloud wind speed equals the environmental wind speed at the top level:
\begin{equation}  \label{eq:uc_downdraft}
\begin{split}
\mathrm{downdraft:} \quad
    &\frac{\partial u_c}{\partial z} = -\epsilon ( u - u_c ), \quad u_c|_{z=H} = u, \\
    &\frac{\partial v_c}{\partial z} = -\epsilon ( v - v_c ), \quad v_c|_{z=H} = v.
\end{split}    
\end{equation}
The above treatment views momentum as a passive tracer. This is inaccurate because momentum is also influenced by the horizontal pressure gradient force \cite[e.g.,][]{lemone1984momentum,wu1994effects}. The horizontal pressure gradient force can be parameterized as a form drag that is approximately proportional to $(u-u_c)$ and $(v-v_c)$, indicating that it is proportional to the direct transport effect \citep{romps2012equivalence}. Therefore, the pressure effect might be parameterized as an effective fractional entrainment rate, which can be included in $\epsilon$. We follow \citet{romps2014rayleigh} to not additionally consider the modification to $\epsilon$ by pressure in this mechanistic analysis, essentially ignoring the pressure effect. 


The momentum carried by the convective plume feedbacks to the environment via the compensating vertical motion and detrainment. Their bulk effect is depicted as: 
\begin{equation} \label{eq:dudt}
\begin{split}
    &\frac{\partial u}{\partial t} =...\,\,+\,\,\frac{\partial}{\partial z} \left[ \frac{M}{\rho} (u-u_c) \right], \\
    &\frac{\partial v}{\partial t} =...\,\,+\,\,\frac{\partial}{\partial z} \left[ \frac{M}{\rho} (v-v_c) \right],
\end{split}    
\end{equation}
where ``..." represents terms in the momentum equation other than the CMT effect, and $\rho$ is treated as a constant quantity in this Boussinesq model (except for the density effect in buoyancy). 


Performing a curl over Eq. (\ref{eq:dudt}), we obtain the expression of CMT in the vorticity equation [Eqs. (\ref{eq:eta}) and (\ref{eq:zeta})]:
\begin{equation} \label{eq:C_expression}
    C_\eta = \frac{\partial^2}{\partial z^2} \left[ \frac{M}{\rho} (u-u_c) \right], 
    \quad
    C_\zeta = \frac{1}{r}\frac{\partial^2}{\partial r \partial z}  \left[ r \frac{M}{\rho} (v-v_c) \right].
\end{equation}
\citet{romps2014rayleigh} showed that the effect of CMT can be understood as a linear damping (Rayleigh damping) on the environmental wind, which is increasingly more significant for a more refined vertical wind structure (higher-order baroclinic modes). The derivation of the Rayleigh damping timescale $\tau_d$ is reviewed in Appendix C. The damping is zero for a barotropic wind mode, because there is nothing to transport in a vertically uniform state. For the first-baroclinic-mode wind structure, i.e., $\zeta \sim \cos \left( \pi z/H \right)$, there is:
\begin{equation}  \label{eq:tau_Rayleigh}
    \mathrm{first\,\,baroclinic\,\,mode: \quad} C_\eta \approx - \frac{\eta}{\tau_d}, \quad
    C_\zeta \approx - \frac{\zeta}{\tau_d}, \quad 
    \tau_d = \frac{\rho}{M}\frac{\epsilon^2 + \pi^2/H^2 }{\epsilon \pi^2/H^2}.
\end{equation}
This derivation only considers the eddy entrainment and detrainment and lets them equal ($\epsilon=\epsilon_{edd}=\delta_{edd}$), so $M$ does not change with height. Equation (\ref{eq:tau_Rayleigh}) shows that a greater convective mass flux amplitude (higher $M_*$) enhances CMT. The $\tau_d$ first decreases with entrainment and then increases with entrainment, contributed by the $\epsilon_{edd}$ in the denominator and numerator, respectively. As explained by \citet{romps2014rayleigh}, in the weak entrainment regime, a stronger entrainment allows the plume momentum ($u_c$, $v_c$) to more efficiently diffuse the environmental momentum from nearby vertical levels, diminishing the vertical wind shear. In the strong entrainment regime, the in-plume wind is too close to the environmental wind, and a stronger entrainment only makes it closer. As a result, the wind difference between the plume and the environment vanishes, making the plume ``transparent" to the environment and transporting momentum less efficiently. A first-baroclinic-mode wind profile in the real atmosphere will be shown to fall within the strong entrainment regime.

In section \ref{sec:simulation}, we perform axisymmetric simulations and show an intriguing result: a barotropic cyclone gradually builds up near the domain center.

\subsection{Experimental parameters}

Simulations are performed in a 512 km $\times$ 10 km cylindrical domain. The initial condition is quiescent, with no wind and buoyancy perturbations. The convective region has a horizontal length scale of $L_M=80$ km, mimicking a mesoscale convective system that could develop into a TC precursor vortex \cite[e.g.,][]{houze2009convective}. A small horizontal and vertical eddy viscosity of $K_h=K_z=2$ m$^2$ s$^{-1}$ is used to stabilize the numerical integration. 
The oscillation period of the convective mass flux uses $T=1$ day, corresponding to the diurnal cycle. The simulation length is five days.

A reference simulation will be analyzed in more detail, with parameters listed in Table \ref{Table_reference}. Other experiments are modified from the reference experiment. We explain the choice of $\epsilon_{edd}$ and $M_*$. The choice of $\epsilon_{edd}=1$ km$^{-1}$ is based on the diagnostic result of \citet{romps2010direct}. The choice of $M_*$ is based on the observational data analysis of Hurricane Karl (2010) by \citet{davis2012mesoscale}. The daily rainfall rate oscillation amplitude at the vortex circulation center is found to be around $\mathcal{R}=5$ mm hour$^{-1}$. Letting the water vapor mixing ratio in the boundary layer be $q_v=10$ g kg$^{-1}$ and the liquid water density be $\rho_l=10^3$ kg m$^{-3}$, we estimate $M_*$ as:
\begin{equation}
    M_* = \frac{\rho_l \mathcal{R}}{q_v} = 0.14\,\,\mathrm{kg\,\,m^{-2}\,\,s^{-1}}.
\end{equation}
For convenience, we use $M_*=0.15$ kg m$^{-2}$ s$^{-1}$ in the reference experiment.

Three groups of sensitivity experiments have been performed, with eight experiments in each group. The varied parameters are summarized in Table \ref{Table_list}. The first group, referred to as CMT\_off\_1-8, changes the oscillation period of the convective mass flux $M_*$ between 0.025-0.2 kg m$^{-2}$ s$^{-1}$, but turns off the convective momentum transfer process (CMT). In other words, convection only influences the system through diabatic heating. The second group, referred to as CMT\_on\_1-8, is similar CMT\_off\_1-8, but includes CMT. The third group, referred to as EDD\_1-8, fixes $M_*$ and changes the eddy fractional entrainment rate $\epsilon_{edd}$ between $0$-1.75 km$^{-1}$. Among these experiments, CMT\_on\_6 and EDD\_5 are identical and taken as the reference experiment. Its no-CMT counterpart, CMT\_off\_6, is an accompanying experiment that is usually mentioned together.

\begin{table}[h]
\caption{The parameters for the reference axisymmetric simulation. The upper panel shows the physical parameters, and the lower panel shows the discretization parameters. }\label{Table_reference}
\begin{center}
\begin{tabular}{cccccrc}
\hline
Symbol & Meaning & Value \\
\hline
$R$  & domain radius & 512 km \\
$H$ & domain height & 10 km \\
$f$ & Coriolis parameter & $5\times10^{-5}$ s$^{-1}$ \\
$\rho$ & air density & 1 kg m$^{-3}$ \\
$M_*$ & amplitude of the convective mass flux & 0.15 kg m$^{-2}$ s$^{-1}$ \\
$L_M$ & horizontal scale of the convective region & 80 km \\
$L_s$ & horizontal scale of the sponge layer & 50 km \\
$\epsilon_{edd}$ & eddy fractional entrainment rate & 1 km$^{-1}$ \\
$\delta_{edd}$ & eddy fractional detrainment rate & 1 km$^{-1}$ \\
$\tau_s$ & sponge damping timescale & 1000 s \\
$N$ & buoyancy frequency & 0.01 s$^{-1}$ \\
$K_h$ & horizontal eddy diffusivity & 2 m$^2$ s$^{-1}$ \\
$K_z$ & vertical eddy diffusivity & 2 m$^2$ s$^{-1}$ \\
\hline 
$N_r$  & radial cell number & 256 \\
$N_z$  & vertical cell number & 40 \\
$\Delta r$  & radial grid spacing & 2 km \\
$\Delta z$  & vertical grid spacing & 0.25 km \\
$\Delta t$  & time step & 20 s \\
\hline
\end{tabular}
\end{center}
\end{table}

\begin{table}[h]
\caption{The experimental list. CMT\_off\_1-8 studies the sensitivity to the mass flux oscillation amplitude $M_*$, without convective momentum transfer. CMT\_on\_1-8 turns on the convective momentum transfer. EDD\_1-8 studies the sensitivity to the eddy fractional entrainment rate $\epsilon_{edd}$. }\label{Table_list}
\begin{center}
\begin{tabular}{cccccrc}
\hline
Name & $M_*$ (kg m$^{-2}$ s$^{-1}$) & $\epsilon_{edd}$ (km$^{-1}$) & CMT \\
\hline
CMT\_off\_1 & 0.025  & 1 & off \\
CMT\_off\_2 & 0.05 & 1 & off \\
CMT\_off\_3 & 0.075 & 1 & off \\
CMT\_off\_4 & 0.1 & 1 & off \\
CMT\_off\_5 & 0.125 & 1 & off \\
CMT\_off\_6 & 0.15 & 1 & off \\
CMT\_off\_7 & 0.175 & 1 & off \\
CMT\_off\_8 & 0.2 & 1 & off \\
\hline
CMT\_on\_1 & 0.025 & 1 & on \\
CMT\_on\_2 & 0.05 & 1 & on \\
CMT\_on\_3 & 0.075 & 1 & on \\
CMT\_on\_4 & 0.1 & 1 & on \\
CMT\_on\_5 & 0.125 & 1 & on \\
CMT\_on\_6 (reference) & 0.15 & 1 & on \\
CMT\_on\_7 & 0.175 & 1 & on \\
CMT\_on\_8 & 0.2 & 1 & on \\
\hline
EDD\_1 & 0.15 & 0 & on \\
EDD\_2 & 0.15 & 0.25 & on \\
EDD\_3 & 0.15 & 0.5 & on \\
EDD\_4 & 0.15 & 0.75 & on \\
EDD\_5 (reference) & 0.15 & 1 & on \\
EDD\_6 & 0.15 & 1.25 & on \\
EDD\_7 & 0.15 & 1.5 & on \\
EDD\_8 & 0.15 & 1.75 & on \\
\hline
\end{tabular}
\end{center}
\end{table}

\section{Axisymmetric simulation results}\label{sec:simulation}

\subsection{The reference experiment}

\begin{figure}[h]
\centerline{\centering
\includegraphics[width=1.0\linewidth]{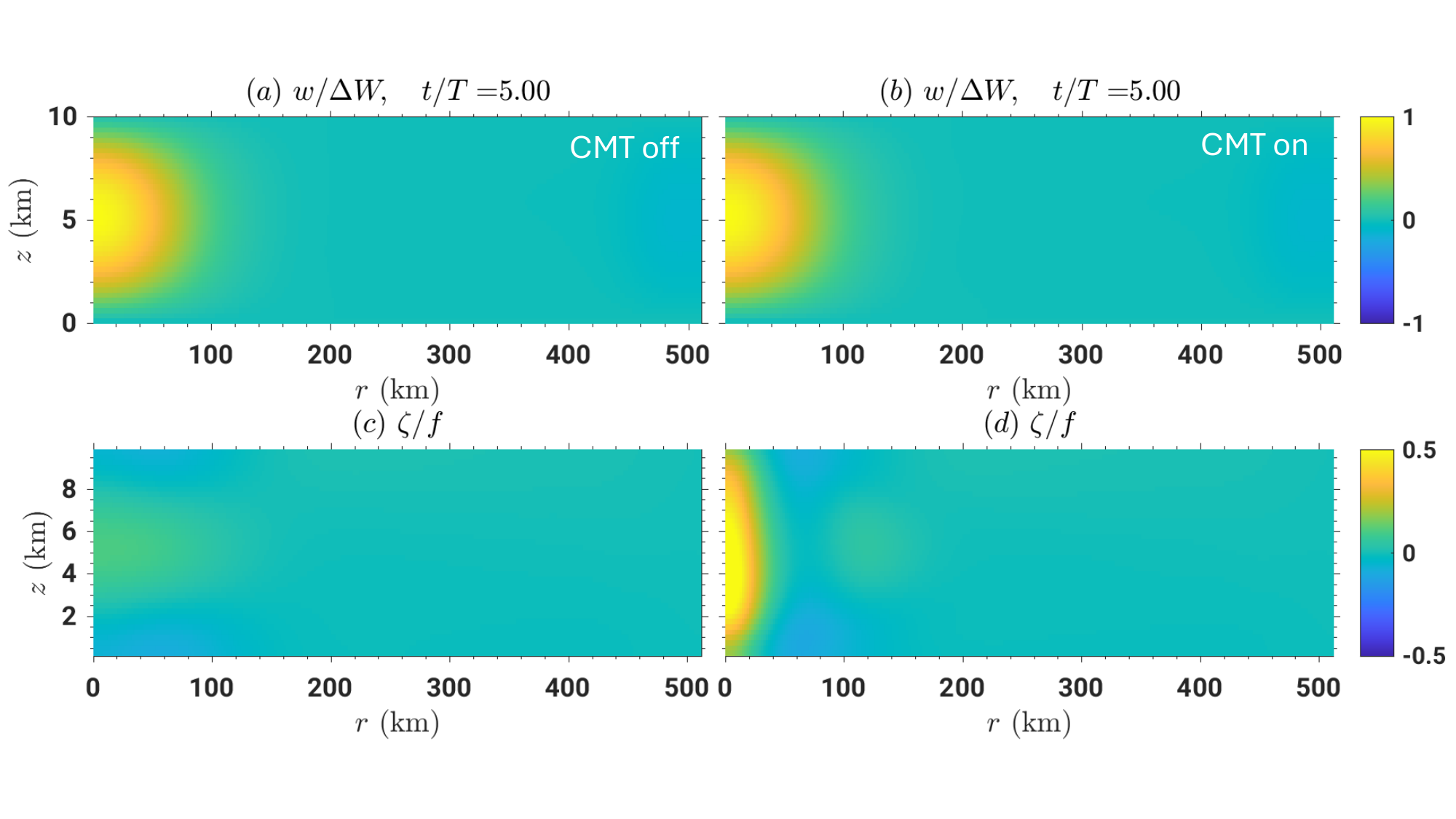}}
  \caption{A snapshot of the reference axisymmetric simulation at $t=5$ days for the reference experiment. The first and second column shows the experiments without and with convective momentum transfer (CMT\_off\_6 and CMT\_on\_6). The first row shows the vertical velocity divided by the characteristic scale of vertical velocity $\Delta W \equiv M_*/\rho$, and the second row shows the vertical vorticity divided by $f$. Movies of all experiments can be downloaded from: https://box.nju.edu.cn/d/653d1189c69c40f8be58/.}\label{fig:2D_display}
\end{figure}

\begin{figure}[h]
\centerline{\centering
\includegraphics[width=1.0\linewidth]{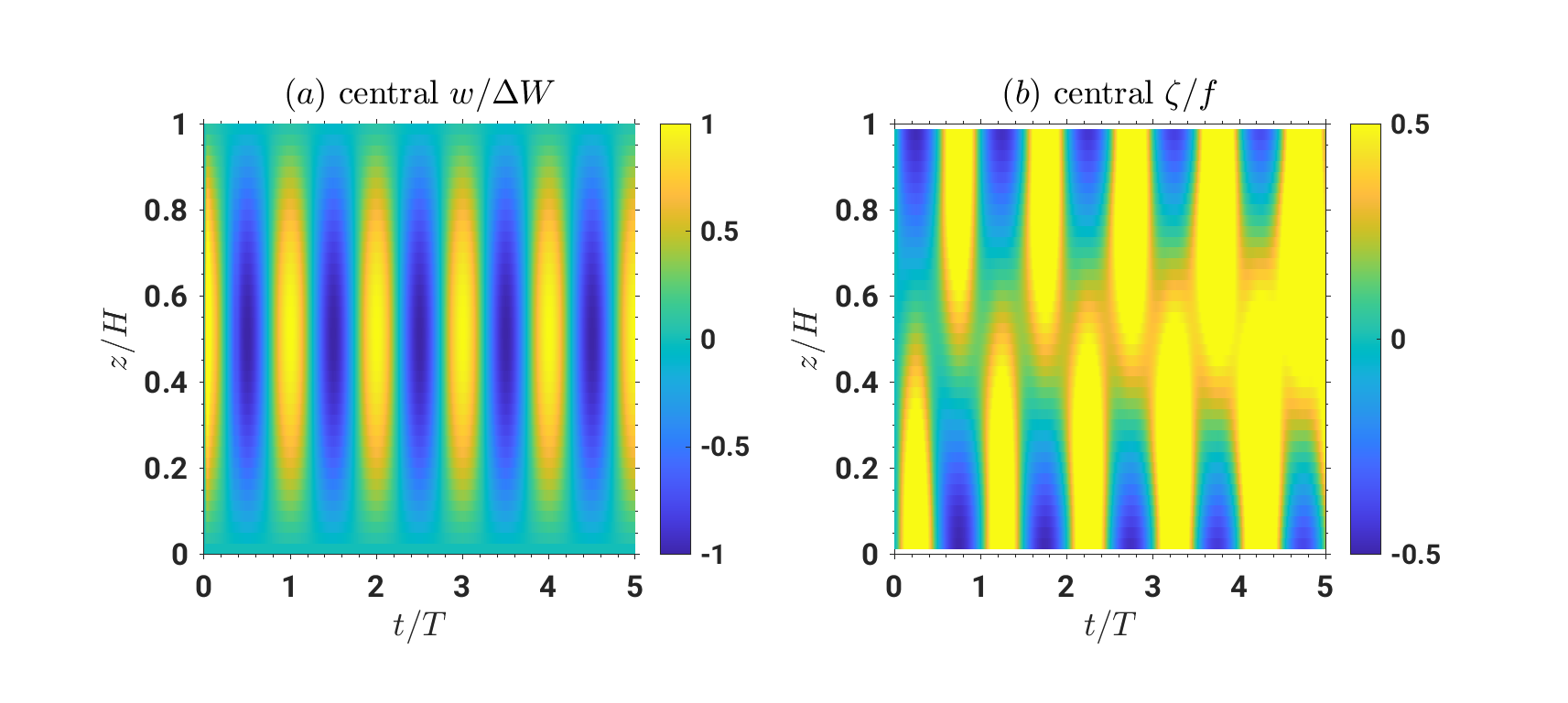}}
  \caption{The Hovmöller diagram of (a) vertical velocity and (b) vertical vorticity at the domain center for the reference experiment (CMT\_on\_6). The horizontal axis is the time $t$ divided by the forcing period $T$, and the vertical axis is the height divided by the domain thickness $H$.}\label{fig:w_Iz_center}
\end{figure}

\begin{figure}[h]
\centerline{\centering
\includegraphics[width=1.1\linewidth]{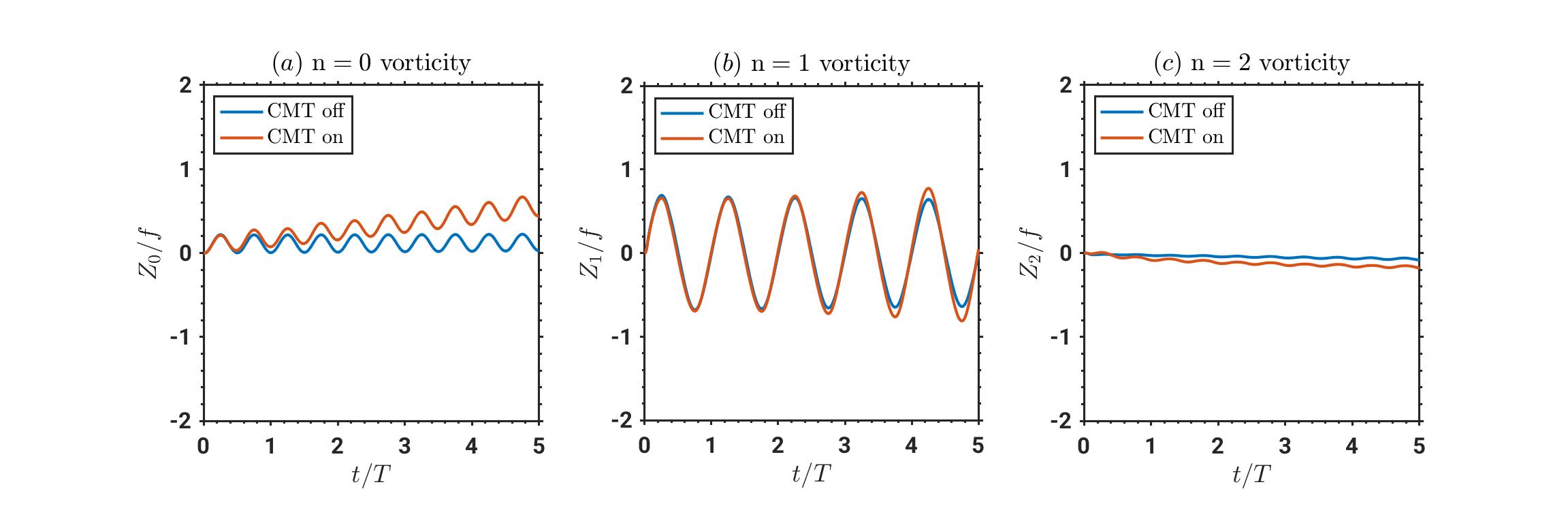}}
  \caption{A vertical mode decomposition analysis of the vertical vorticity at the domain center, using the reference experiment (CMT\_on\_6) and its counterpart (CMT\_off\_6) that turns off the CMT. (a) The barotropic vorticity $Z_0/f$. (b) The first baroclinic mode vorticity $Z_1/f$. (c) The second baroclinic mode vorticity $Z_2/f$. }\label{fig:Z0_Z1_Z2}
\end{figure}

Figure \ref{fig:2D_display} shows the distribution of vertical velocity $w$ and vorticity $\zeta$ at $t=5$ days, using the reference experiment. Comparing the cases without and with convective momentum transfer, the $w$ is not significantly different. However, $\zeta$ has a deep (vertically uniform) cyclonic core at the domain enter and a deep anticyclonic shell around it. To check whether the deep vorticity anomaly is a transient feature of this snapshot, Fig. \ref{fig:w_Iz_center} shows the Hovmöller diagram of $\zeta$ and $w$ at the domain center. The $w$ responds to the convective activity in an oscillatory way, with a steady amplitude. The $\zeta$ also responds in an oscillatory way, but a deep vorticity anomaly steadily grows, confirming the visual inspection of Fig. \ref{fig:2D_display}. To further quantify the vorticity structure at the vortex center, we perform a vertical mode decomposition \cite[e.g.,][]{vallis2017atmospheric}:
\begin{equation}
\begin{split}
   Z_n \equiv \begin{cases}
       \frac{2}{H} \int_0^H \zeta \cos\left( \frac{n \pi z}{H} \right) dz, \quad n=0,\\
       \frac{1}{H} \int_0^H \zeta \cos\left( \frac{n \pi z}{H} \right) dz, \quad n \ge 1,       
   \end{cases} \quad \zeta = \sum_{n=0}^\infty Z_n \cos \left( \frac{n \pi z}{H} \right).
\end{split}   
\end{equation}
The $Z_n$ is the amplitude of each vertical mode of vertical vorticity, with $Z_0 = \overline{\zeta}$ representing the barotropic mode, and the overbar denoting the vertical average. Figure \ref{fig:Z0_Z1_Z2} shows that when CMT is active, the barotropic vorticity grows steadily, superposed with small ripples of half the forcing period. The first-baroclinic-mode vorticity $Z_1$ oscillates with the forcing period, with the amplitude gradually growing. The $Z_2$ vorticity is negligible. Three key results are summarized below:
\begin{enumerate}
    \item A barotropic cyclone grows at the domain center.
    \item A barotropic anticyclonic shell wraps the barotropic cyclone.    
    \item The amplitude of the $n=1$ vorticity grows at the domain center.     
\end{enumerate}
To explain them, we analyze the sensitivity experiments that change $M_*$ and $\epsilon_{edd}$. 

\subsection{Sensitivity experiments}

Sensitivity experiments that change $M_*$ and $\epsilon_{edd}$ are analyzed, focusing on variables at the domain-center. 

\begin{figure}[h]
\centerline{\centering
\includegraphics[width=1.0\linewidth]{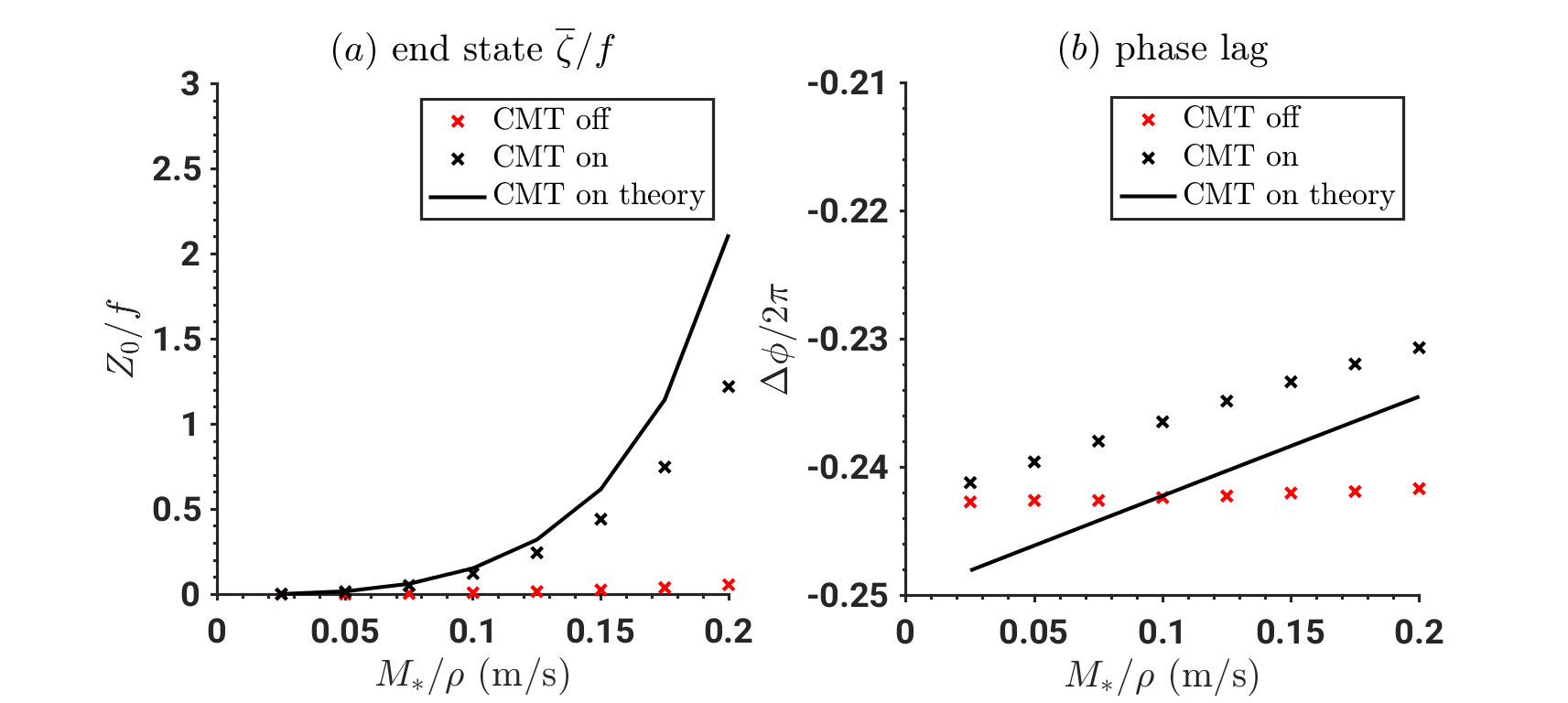}}
  \caption{Results of the first two groups of experiments (CMT\_off and CMT\_on). (a) The barotropic vorticity at the domain center at $t=5$ days for experiments with different $M_*$. The red crosses show experimental results where CMT is turned off; the black crosses show the experiments where CMT is active. The black line shows the theoretical prediction when CMT is active. (b) is the same as (a) but for the phase lag of $Z_1$ to $w(z=H/2)$. The diagnosed phase lag has a ``background level" that makes the lag slightly smaller than $\pi/2$. This background level is likely a technical issue in calculating autocorrelation due to the finite length of the data, and is hard to eliminate.    }\label{fig:lag_w0}
\end{figure}

\begin{figure}[h]
\centerline{\centering
\includegraphics[width=1.0\linewidth]{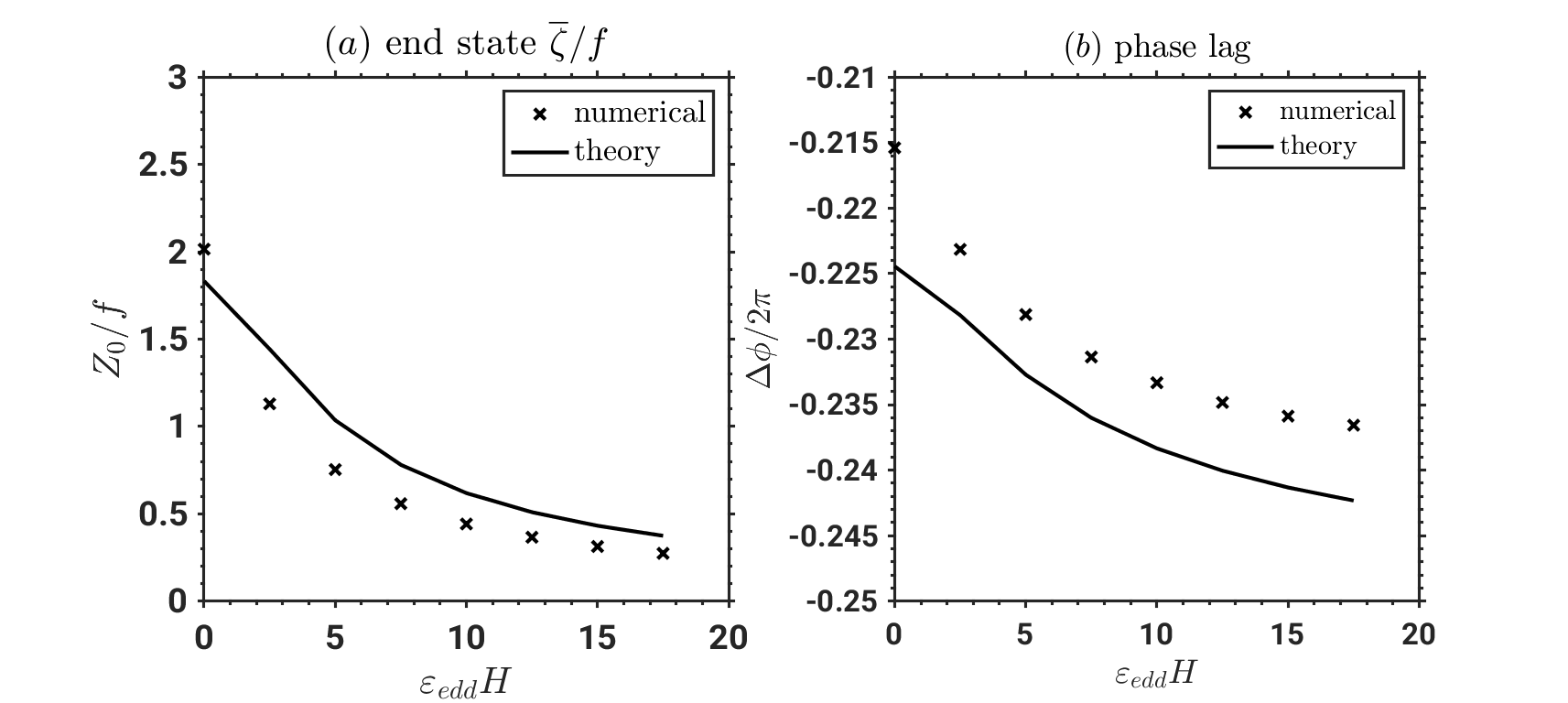}}
  \caption{Results of the third group of experiments (EDD). (a) The barotropic vorticity at the domain center at $t=5$ days for experiments with different $\epsilon_{edd}$. The black crosses show the experimental results. The black line shows the theoretical prediction. (b) is the same as (a) but for the phase lag of $Z_1$ to $w(z=H/2)$. }\label{fig:lag_entrain}
\end{figure}

Comparing the first two groups (CMT\_off) and (CMT\_on) where $M_*$ is changed, we confirm that CMT is essential for the growth of barotropic vorticity, and the end-state barotropic vorticity increases significantly with $M_*$ (Fig. \ref{fig:lag_w0}a). Then, we analyze the EDD experiments with different $\epsilon_{edd}$. Both the dynamical entrainment and eddy entrainment contribute to the entrainment rate. Thus, even if $\epsilon_{edd}=0$, the CMT due to the dynamical entrainment still exists. Figure \ref{fig:lag_entrain} shows that a greater $\epsilon_{edd}$ produces less barotropic vorticity. This, together with Fig. \ref{fig:lag_w0}, indicates that CMT is essential for the production of a barotropic cyclone, but too strong an entrainment reduces the production rate.

\section{Theory}\label{sec:theory}

\subsection{Dynamics at the domain center}

We adopt a variable-separation approach to analyze the production of barotropic vorticity. The main body text focuses on the vertical structure, characterized by the vorticity dynamics at the domain center. Appendix D studies the radial structure.  

\begin{figure}[h]
\centerline{\centering
\includegraphics[width=0.5\linewidth]{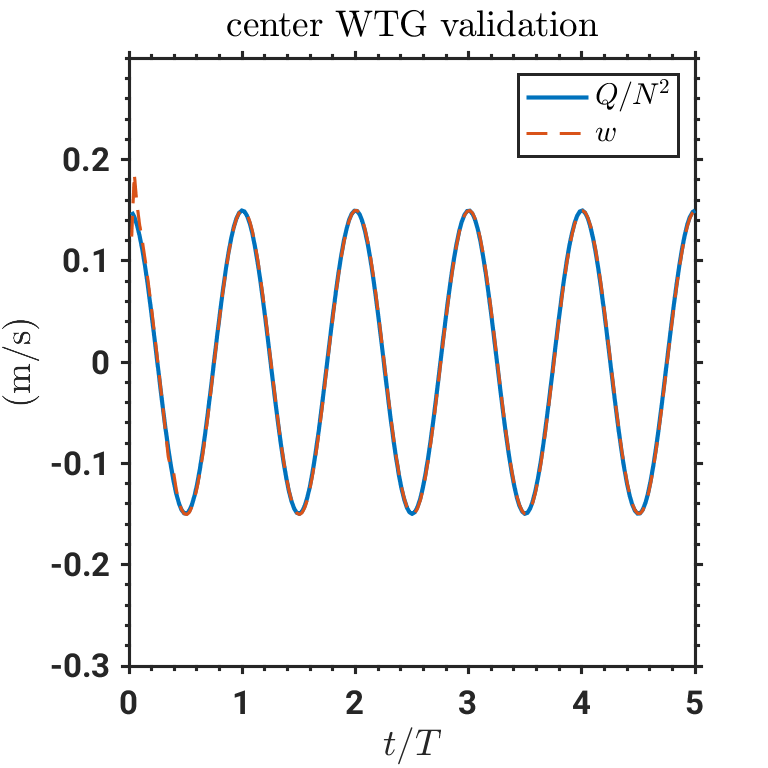}}
  \caption{A validation of the weak temperature gradient approximation at the domain center [Eq. (\ref{eq:w_WTG})], using the reference experiment (CMT\_on\_6). The solid blue line shows $Q/N^2$, and the dashed red line shows $w$. }\label{fig:WTG}
\end{figure}

At the domain center, there is $u=v=0$, and $\partial w/\partial r = \partial \zeta/\partial r = \partial M/\partial r = 0$. The $\zeta$ equation reduces to:
\begin{equation}  \label{eq:zeta_center}
    \frac{\partial \zeta}{\partial t} = - w\frac{\partial \zeta}{\partial z} 
    + \left( f + \zeta \right) \frac{\partial w}{\partial z} + \frac{\partial}{\partial z} \left[ \frac{M}{\rho} \left( \zeta - \zeta_c \right) \right],
\end{equation}
where the CMT expression is derived from Eq. (\ref{eq:C_expression}). The in-cloud vorticity equation is derived from Eqs. (\ref{eq:uc_updraft}) and (\ref{eq:uc_downdraft}):
\begin{equation}  \label{eq:zeta_c}
    \frac{\partial \zeta_c}{\partial z} = 
    \begin{cases}
        \epsilon (\zeta - \zeta_c), \quad \mathrm{updraft}, \\
        -\epsilon (\zeta - \zeta_c), \quad \mathrm{downdraft}.
    \end{cases}
\end{equation}
The grid-scale vertical velocity $w$ can be approximately calculated with the weak temperature gradient approximation \cite[WTG,][]{sobel2001WTG}:
\begin{equation}  \label{eq:w_WTG}
    w = \frac{Q}{N^2} = \frac{M_*}{\rho},
\end{equation}
which is reduced from the buoyancy equation (\ref{eq:b}) by letting the adiabatic cooling term directly balance the diabatic heating term. This approximation is valid when 1) the convective region is narrow, 2) the convective oscillation is slow, and 3) the vorticity is weak. By narrow, we mean the convective-region length scale ($L_M$) is much smaller than the Rossby deformation radius ($NH/\pi f$) near the vortex center. This results in the compensating descent occurring in a much broader area than the convective region and therefore being negligible at the domain center \citep{sobel2001WTG}. By slow, we mean that the oscillatory frequency of convective activity ($\Omega$) is not significantly longer than the timescale of geostrophic adjustment, $f^{-1}$ \cite[e.g.,][]{gill1982atmosphere}, which is the time required for compensating descent to develop. By weak, we mean the relative vorticity ($\zeta$) does not significantly influence the inertial stability parameter \cite[e.g.,][]{hack1986nonlinear}. Based on the scale analysis of \citet{sobel2001WTG}, these three conditions are quantified as:
\begin{equation}  \label{eq:WTG_condition}
    \max\left\{ \frac{1}{fT},\,\frac{\zeta}{f} \right\} \,\mathrm{Bu} \ll 1,\quad \mathrm{where} \quad \mathrm{Bu} \equiv \frac{L_M}{NH/\pi f}.
\end{equation}
Here, Bu is the Burger number, which represents ``how small"; $1/fT$ is the Kibel number, which represents ``how slow"; $\zeta/f$ is the Rossby number, which represents ``how weak". Substituting in the parameters in Table \ref{Table_reference}, and assuming $\zeta/ f\to 0$, the left-hand side of Eq. (\ref{eq:WTG_condition}) is 0.0037, indicating that WTG is well satisfied for an infinitesimally weak vortex. To make the left-hand side term reach 0.1 and violate WTG, $\zeta/f$ should reach 6, which is beyond all our experiments (see Figs. \ref{fig:lag_w0}a and \ref{fig:lag_entrain}a). In addition to the three conditions, the Rayleigh damping posts another requirement:
\begin{equation}
    \Omega \tau_d \ll 1.
\end{equation}
Otherwise, the Rayleigh damping (CMT) would significantly damp the overturning circulation and reduce $w$. The applicability of WTG approximation is validated by the simulation (Fig. \ref{fig:WTG}).\footnote{\citet{robinson2008resonant} proposed that a periodic wave source from low-level heating with finite horizontal width could excite ``trapped" gravity wave at the origin, which may resonate with convection. \citet{yang2024internal} showed that a transient internal diabatic heating can generate a decaying oscillation that couples with convection. Nevertheless, our simulation shows that the WTG works well at the domain center, indicating the response is largely balanced rather than wave-like. The discrepancy may be due to the different sampling regimes in the parameter space \citep{navarro2017balanced}, and further investigation is needed.  }

Equations (\ref{eq:zeta_center})-(\ref{eq:w_WTG}), together with the expression of $M$ [Eq. (\ref{eq:M_structure})] and entrainment rate $\epsilon$ [Eqs. (\ref{eq:epson_sum}), (\ref{eq:epson_delta_edd}) (\ref{eq:dyn_updraft}), (\ref{eq:dyn_downdraft})], constitute a 1D column model of vorticity dynamics at the domain center. The main difference from the single-column model of \citet{fu2025role} is the inclusion of a bulk-plume convective model and the change from steady to oscillatory forcing.

\subsection{Key physics: CMT induces a phase shift}\label{subsec:physics}

Next, we use a heuristic asymptotic argument to understand the production mechanism of barotropic vorticity, followed by a more rigorous analysis. Imposing a vertical average over Eq. (\ref{eq:zeta_center}), we obtain:
\begin{equation}  \label{eq:zeta_mean_center}
    \frac{\partial \overline{\zeta}}{\partial t} = - \overline{w\frac{\partial \zeta}{\partial z}} 
    + \overline{\zeta \frac{\partial w}{\partial z}}.
\end{equation}
Here, $\overline{f\partial w / \partial z}$ vanishes because $w=0$ at the domain bottom and top. We deliberately remove CMT in the analysis, which reveals why it is indispensable for the production of barotropic vorticity.

When $w$ is steady with time, \citet{fu2025role} showed that the two right-hand side terms of Eq. (\ref{eq:zeta_mean_center}) both contribute to the barotropic vorticity. To see this, suppose $\zeta$ can be approximated as its linear response component to $w \sim \sin \left( \pi z/H \right)$:
\begin{equation}  \label{eq:vort_linear}
    \frac{\partial \zeta}{\partial z} \sim f \frac{\partial w}{\partial z},
\end{equation}
which yields $\zeta \sim Z_1 \cos\left( \pi z/H \right)$. The sum of vertical advection and stretching terms obeys:
\begin{equation}
\mathrm{steady:}\quad
     - \overline{w\frac{\partial \zeta}{\partial z}} 
    + \overline{\zeta \frac{\partial w}{\partial z}}
    \sim \overline{\sin^2 \left( \frac{\pi z}{H} \right)} + \overline{\cos^2 \left( \frac{\pi z}{H} \right)},
\end{equation}
which has a vertically uniform structure. 

When $w$ is periodic in time, as in this problem, the result differs. Letting $w \sim \sin \left( \pi z/H \right) \cos \left( \Omega t \right)$, we use Eq. (\ref{eq:vort_linear}) to obtain $\zeta \sim - \cos\left( \pi z/H \right) \sin \left( \Omega t \right)$. The sum of the two terms, averaged over a period $T$, obeys:
\begin{equation}
\begin{split}
\mathrm{oscillatory:}\quad
    &\frac{1}{T} \int_0^T \left( - \overline{w\frac{\partial \zeta}{\partial z}  } 
    + \overline{\zeta \frac{\partial w}{\partial z}} \right) dt \\
    &\sim - \frac{1}{T} \int_0^T \overline{\sin^2 \left( \frac{\pi z}{H} \right) \sin(\Omega t) \cos(\Omega t)} dt - \frac{1}{T} \int_0^T \overline{\cos^2 \left( \frac{\pi z}{H} \right) \sin(\Omega t) \cos(\Omega t)} dt \\
    &=0.
\end{split}    
\end{equation}
Thus, to make the sum of the two terms non-zero, $w$ and the linear response component of $\zeta$ should not be in quadrature, i.e., their phase lag should not equal to $\pi/2$. 

To confirm this heuristic argument, we analyze the phase relation of $w$ (at $z=H/2$) and $Z_1$ (Fig. \ref{fig:lag_w0}b). The phase lag is diagnosed by calculating the cross-correlation coefficient between $w(z=H/2)$ and $Z_1$ spanning $t=1$ days and $t=5$ days. The time lag where the correlation coefficient peaks is used to calculate the phase lag. Without CMT, $Z_1$ and $w(z=H/2)$ are in quadrature, with $Z_1$ lagging $w(z=H/2)$ by approximately $\pi/2$. With CMT, the phase lag drops, and the drop becomes more significant as $M_*$ increases. This analysis indicates that the production of barotropic vorticity is associated with the CMT's modulation on the phase relation between $w$ and $\zeta$. The modulation mechanism will be more rigorously analyzed in section \ref{sec:theory}\ref{subsec:reduced_model}.

\subsection{A reduced model} \label{subsec:reduced_model}

The single-column model, which depicts the dynamics at the vortex center, can be further simplified to a model with an analytical solution.

\subsubsection{Linearizing the CMT term}

The key assumption is linearizing the CMT term, making $\tau_d$ a constant quantity in both space and time. According to Eq. (\ref{eq:tau_Rayleigh}), the key is approximating $M$ to be invariant with time and height.

Making $\tau_d$ steady with time requires $M$ to be steady with time. However, $M$ is oscillatory in this problem, with $M$ taking $M_*$ at the peak of the updraft and downdraft phase and taking $0$ at the transitional phase. Therefore, we use $M\approx M_*/2$ to represent the $M$ in the $\tau_d$ expression (\ref{eq:tau_Rayleigh}).

Making $\tau_d$ uniform with height requires $M$ to be uniform with height. This requires $\epsilon=\delta$, which indicates that there is no dynamical entrainment and detrainment, and the eddy entrainment and detrainment rates are equal ($\epsilon_{edd}=\delta_{edd}$). While $\epsilon_{edd}=\delta_{edd}=\mathrm{const}$ is a reasonable approximation in analyzing the CMT only \citep{romps2014rayleigh}, neglecting dynamical entrainment means there is no diabatic heating, which is inconsistent with the prescribed sinusoidal vertical structure of $M$ [Eq. (\ref{eq:M_structure})]. To address the gap, we approximate $\epsilon_{dyn}$ as the vertically averaged value in calculating $\tau_d$ [Eq. (\ref{eq:tau_Rayleigh})]:
\begin{equation} \label{eq:tao_d_modified}
    \tau_d \approx \frac{\rho}{M_*/2}\frac{(\epsilon_{edd} + \overline{\epsilon_{dyn}})^2 + \pi^2/H^2 }{(\epsilon_{edd}+\overline{\epsilon_{dyn}}) \pi^2/H^2},
    \quad \mathrm{where} \quad
    \overline{\epsilon_{dyn}} \equiv \frac{1}{H}\int_0^H \epsilon_{dyn} dz.
\end{equation}
Here, $\epsilon_{dyn}$ obeys a piecewise cotangent function in $z$ [Eqs. (\ref{eq:dyn_updraft}) and (\ref{eq:dyn_downdraft})]. Unfortunately, at $z=0$ or $z=H$, the cotangent function is unbounded. This makes the value of $\overline{\epsilon_{dyn}}$ unbounded. In the numerical model, this indicates that $\overline{\epsilon_{dyn}}$ depends on the vertical grid spacing $\Delta z$.


Physically, the origin of this uncertainty is the prescribed sinusoidal vertical structure of $M$, which is zero at the bottom and top. In the real atmosphere, the convective mass flux is defined above the mixed layer top, and there must be a non-zero cloud-base mass flux that serves as the boundary condition for the $M$ equation (\ref{eq:M_epsilon}). Such a boundary condition can be provided by various closure assumptions, such as the convective quasi-equilibrium closure \citep{arakawa1974interaction,pan1998cumulus} or the gust front lifting closure \citep{grandpeix2010density}. 

\begin{figure}[h]
\centerline{\centering
\includegraphics[width=1.15\linewidth]{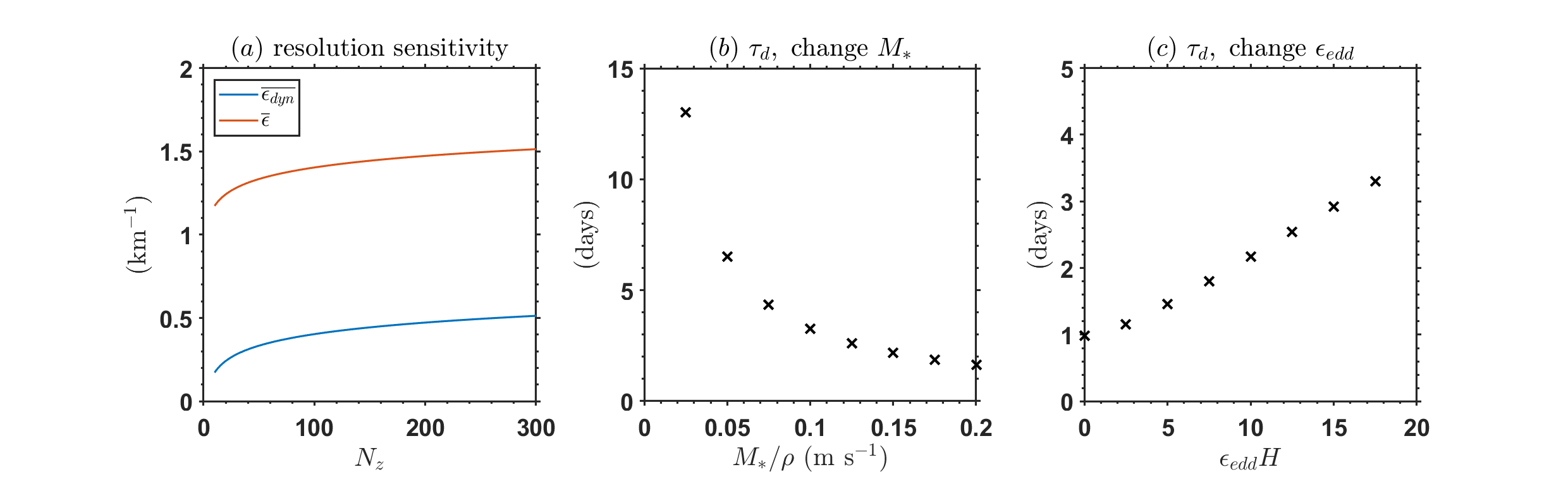}}
  \caption{(a) The sensitivity of $\overline{\epsilon_{dyn}}$ (blue line) and $\overline{\epsilon}$ (red line) to the vertical cell number $N_z$. The calculation uses Eq. (\ref{eq:epson_dyn_numerical}) and $\overline{\epsilon} = \overline{\epsilon_{dyn}} + \epsilon_{edd}$. (b) The dependence of the $n=1$ mode Rayleigh damping timescale, $\tau_d$, on the convective mass flux amplitude $M_*$. (c) The dependence of $\tau_d$ on the eddy fractional entrainment rate $\epsilon_{edd}$.}\label{fig:Nz}
\end{figure}

To be compatible with the axisymmetric simulation, we temporarily retain this uncertainty and use the numerically calculated $\epsilon_{dyn}$ in the theoretical model. The $\epsilon_{dyn}$ is defined at $z=\Delta z,\,\,2\Delta z, ..., H-\Delta z$. Taking the updraft phase as an example, $\overline{\epsilon_{dyn}}$ is calculated with:
\begin{equation}  \label{eq:epson_dyn_numerical}
   \overline{\epsilon_{dyn}} \equiv \frac{1}{N_z-1}\sum_{k=2}^{Nz/2} \frac{\pi}{H} \cot \left[ \frac{\pi}{H} \Delta z (k-1) \right].   
\end{equation}
The downdraft phase has the same result. At $z=0$, $\epsilon_{dyn}$ is not defined because it is the initiation height of updraft and downdraft, where $u=u_c$ and $v=v_c$ [Eqs. (\ref{eq:uc_updraft}) and (\ref{eq:uc_downdraft})]. This waives the requirement of including the infinite $\cot(0)$ value at $z=0$. Figure \ref{fig:Nz} shows the numerical calculation result with $N_z$ changing from 10 to 300. The $\overline{\epsilon_{dyn}}$ increases with $N_z$, but the slope gets flatter after $N_z$ reaches 50 and approximately takes 0.5 km$^{-1}$. The $\epsilon_{edd}$ has been diagnosed to be around 1 km$^{-1}$ by \citet{romps2010direct}. Their sum, $\overline{\epsilon} \equiv \overline{\epsilon_{dyn}}+\epsilon_{edd}$, is even less sensitive to $N_z$. Thus, the CMT in this model is indeed resolution-dependent, but does not significantly influence the result. 

Figure \ref{fig:Nz}b and c show the $\tau_d$ of the eight CMT\_on and EDD experiments calculated with Eqs. (\ref{eq:tao_d_modified}) and (\ref{eq:epson_dyn_numerical}). The $\tau_d$ decreases with $M_*$, because a greater $M_*$ induces stronger vertical transport. The $\tau_d$ increases with $\epsilon_{edd}$, indicating that it reduces the damping. To understand this, we note that a greater $\epsilon_{edd}$ induces stronger lateral transport between the plume and the environment. The experiments are in the strong-entrainment regime where the in-cloud momentum is very close to the environmental value, making the clouds ``transparent" to the wind. Thus, in this regime, a stronger entrainment reduces the damping \citep{romps2014rayleigh}.

\subsubsection{Vertical mode truncation}

With the above preparations, we decompose the single-column vorticity equation (\ref{eq:zeta_center}) into vertical modes and make a truncation that only retains the $n=0$ and $n=1$ modes:
\begin{equation}  \label{eq:zeta_decompose}
 \zeta \approx Z_0 + Z_1 \cos \left( \frac{\pi z}{H} \right).   
\end{equation}
The $n=2$ mode is neglected, given the diagnostic result shown in Fig. \ref{fig:Z0_Z1_Z2}. Substituting Eq. (\ref{eq:zeta_decompose}) into Eq. (\ref{eq:zeta_center}), we get:
\begin{equation}  \label{eq:Z0_oscillatory}
    \frac{dZ_0}{dt}
    = \cos \left( \Omega t \right) \frac{Z_1}{\tau_*},
\end{equation}
\begin{equation}
\label{eq:Z2_oscillatory}
    \frac{dZ_1}{dt}
    = \cos \left( \Omega t \right) \frac{f+Z_0}{\tau_*} - \frac{Z_1}{\tau_d},
\end{equation}
where $\tau_*$ is the overturning timescale:
\begin{equation}
    \tau_* \equiv \left( \frac{M_*}{\rho}\frac{\pi}{H} \right)^{-1}.
\end{equation}
Note that CMT does not act on the $n=0$ mode vorticity. 

\subsubsection{Insights from the reduced model}

The reduced equation [(\ref{eq:Z0_oscillatory}) and (\ref{eq:Z2_oscillatory})] reveals rich physics, even before the analytical solution is derived.

First, the reduced equation reveals how CMT modulates the phase lag between $w$ and $Z_1$. Without CMT, the $\cos\left( \Omega t \right)$ forcing produces $Z_1 \propto \sin \left( \Omega t \right)$. The right-hand side of Eq. (\ref{eq:Z0_oscillatory}) is proportional to $\cos\left( \Omega t \right) \sin \left( \Omega t \right) = \frac{1}{2}\sin \left( 2 \Omega t \right)$, producing a purely oscillatory $Z_0$. With CMT, the lag of $Z_1$ with respect to $w$ decreases. In the limit of $\Omega \tau_d \to 0$, the phase lag vanishes. As a result, the $w$ and $Z_1$ are no longer in quadrature. They produce an increment of $Z_0$ via the vertical advection and stretching of vorticity after each surge of updraft and downdraft. Though a finite CMT is needed to produce an increment, overly damping also suppresses the increment by reducing the $Z_1$ amplitude. Letting $f+Z_0 \approx f$ in Eq. (\ref{eq:Z2_oscillatory}), we can quantitatively estimate the phase lag. Suppose
\begin{equation}  \label{eq:Z1_hat_definition}
  Z_1 = \hat{Z}_1 \exp(i\Omega t), \quad w = \hat{w} \exp(i\Omega t),   
\end{equation}
where $\hat{Z}_1$ is a complex amplitude, and $\hat{w}$ is a real amplitude due to the property of $w \propto \cos(\Omega t)$. Substitute Eq. (\ref{eq:Z1_hat_definition}) into the $Z_1$ equation (\ref{eq:Z2_oscillatory}), we obtain the expression of $\hat{Z}_1$:
\begin{equation}  \label{eq:Z1_complex}
    \hat{Z}_1 = \frac{f \tau^{-1}_*}{\tau^{-1}_d+i\Omega}.
\end{equation}
The phase lag $\Delta \phi$ is calculated with:
\begin{equation} \label{eq:phase_lag}
   \Delta \phi \equiv \mathrm{Arg} \left( \frac{\hat{Z}_1}{\hat{w}} \right) = \mathrm{Arg} \left( \frac{1}{\tau_d} - i\Omega \right),  
\end{equation}
where Arg denotes calculating the phase angle. Figures \ref{fig:lag_w0}b and \ref{fig:lag_entrain}b show that the theoretical prediction of the phase lag [Eq. (\ref{eq:phase_lag})] captures the trend of the axisymmetric simulations.

Second, the reduced equation explains why $Z_1$ amplifies with time (Fig. \ref{fig:Z0_Z1_Z2}b). This is because the growth of $Z_0$ feedbacks to $Z_1$ by enhancing the inertial stability, as shown in the $(f+Z_0)$ term in Eq. (\ref{eq:Z2_oscillatory}). This makes diabatic heating more efficient in generating the $n=1$ vorticity, causing its amplification.

\subsection{The analytical solution}

Next, we solve the growth rate of the barotropic vorticity $Z_0$, which allows us to explore the parameter space. The mathematical technique we adopt is separating the fast- and slow-scale motions, inspired by the parametric resonance problem in the textbook of \citet{landau1960mechanics}. Suppose $Z_0$ and $Z_1$ have the following form:
\begin{equation} \label{eq:special_solution_oscillatory}
\begin{split}
\begin{cases} 
 Z_0 = \widetilde{Z_0}(t)\, + \,\hat{Z}_{0+}(t) \, e^{i2\Omega t} \, + \, \hat{Z}_{0-}(t) \, e^{-i2\Omega t}, \\
Z_1 = \hat{Z}_{1+}(t) \, e^{i\Omega t} \, + \, \hat{Z}_{1-}(t) \, e^{-i\Omega t}.
\end{cases}
\end{split}
\end{equation}
Here $\widetilde{Z_0}(t)$, $\hat{Z}_{0+}(t)$, $\hat{Z}_{0-}(t)$, $\hat{Z}_{1+}(t)$, and $\hat{Z}_{1-}(t)$ are slow-varying functions of time that evolve much slower than the oscillatory convective activity. Substituting Eq. (\ref{eq:special_solution_oscillatory}) into Eqs. (\ref{eq:Z0_oscillatory}) and (\ref{eq:Z2_oscillatory}), we get:
\begin{equation}  \label{eq:Z0_oscillatory_expanded}
\begin{split}
    \frac{d \widetilde{Z_0}}{dt}
    &+ \frac{d\hat{Z}_{0+}}{dt} e^{i2\Omega t}
    + \hat{Z}_{0+} 2i \Omega e^{i2\Omega t}
    + \frac{d\hat{Z}_{0-}}{dt} e^{-i2\Omega t}
    - \hat{Z}_{0-} 2i \Omega e^{-i2\Omega t}\\
    &= \frac{1}{\tau_*} \frac{1}{2} \left( 
    \hat{Z}_{1+} 
    + \hat{Z}_{1-} 
    + \hat{Z}_{1+} e^{i 2\Omega t}
    + \hat{Z}_{1-} e^{-i 2\Omega t}    
    \right),
\end{split}    
\end{equation}
\begin{equation}
\label{eq:Z2_oscillatory_expanded}
\begin{split}
    \frac{d\hat{Z}_{1+}}{dt} e^{i\Omega t}
    &+ \hat{Z}_{1+} i \Omega e^{i\Omega t}
    + \frac{d\hat{Z}_{1-}}{dt} e^{-i\Omega t}
    - \hat{Z}_{1-} i \Omega e^{-i\Omega t}\\
    &= \frac{1}{\tau_*} 
    \frac{e^{i\Omega t} + e^{-i\Omega t}}{2}
    \underbrace{ \left( f + \widetilde{Z_0} + \hat{Z}_{0+} e^{i2\Omega t}  +  \hat{Z}_{0-} e^{-i2\Omega t} \right) }_{\text{inertial stability}}
    - \frac{\hat{Z}_{1+}e^{i\Omega t}}{\tau_d}
    - \frac{\hat{Z}_{1-}e^{-i\Omega t}}{\tau_d}.    
\end{split}    
\end{equation}
Then, two approximations are made:
\begin{itemize}
    \item We ignore the oscillatory component of $Z_0$ in modulating the inertial stability, whose amplitude is small compared to the steady growth component $\widetilde{Z_0}$ (Fig. \ref{fig:Z0_Z1_Z2}a):
\begin{equation}
    f + \widetilde{Z_0} + \hat{Z}_{0+} e^{i2\Omega t}  +  \hat{Z}_{0-} e^{-i2\Omega t} 
    \approx 
     f + \widetilde{Z_0}.
\end{equation}
    \item The slow-varying assumption for the amplitude envelope of $Z_1$ yields:
    \begin{equation}
        O \left( \frac{d\hat{Z}_{1+}}{dt} \right)
         \ll 
        O \left( \hat{Z}_{1+} i \Omega \right)
        ,
        \quad
        O \left( \frac{d\hat{Z}_{1-}}{dt} \right)
        \ll 
        O \left( \hat{Z}_{1-} i \Omega \right).
    \end{equation}
\end{itemize}
Applying the two assumptions and grouping the terms in Eqs. (\ref{eq:Z0_oscillatory_expanded}) and (\ref{eq:Z2_oscillatory_expanded}) by frequency, we get the equations of $\widetilde{Z_0}$, $\hat{Z}_{0+}$, $\hat{Z}_{0-}$, $\hat{Z}_{1+}$, and $\hat{Z}_{1-}$:
\begin{equation} \label{eq:Z0_equation}
    \frac{d \widetilde{Z_0}}{dt}
    =  \frac{1}{2\tau_*}  \left( \hat{Z}_{1+} + \hat{Z}_{1-} \right),
\end{equation}
\begin{equation} \label{eq:Z0+_equation}
    \frac{d \hat{Z}_{0+}}{dt}
    =  \frac{1}{2\tau_*} 
    \hat{Z}_{1+},
\end{equation}
\begin{equation} \label{eq:Z0-_equation}
    \frac{d \hat{Z}_{0-}}{dt}
    =  \frac{1}{2\tau_*} 
    \hat{Z}_{1-},
\end{equation}
\begin{equation} \label{eq:Z2+_equation}
    \hat{Z}_{1+} = \frac{1}{2\tau_*} \frac{ f + \widetilde{Z_0} }{ i \Omega + \frac{1}{\tau_d} },
\end{equation}
\begin{equation} \label{eq:Z2-_equation}
    \hat{Z}_{1-} = \frac{1}{2\tau_*} \frac{ f + \widetilde{Z_0} }{ -i \Omega + \frac{1}{\tau_d} }.
\end{equation}

We are most concerned with $\widetilde{Z_0}$, the steady growth component of the barotropic vorticity. Substituting Eqs. (\ref{eq:Z2+_equation}) and (\ref{eq:Z2-_equation}) into Eq. (\ref{eq:Z0_equation}), we get:
\begin{equation}  \label{eq:Z0_relaxation_equation}
    \frac{d \widetilde{Z_0}}{dt}
    = \frac{f + \widetilde{Z_0}}{\tau_{Z0}},
\end{equation}
where $\tau_{Z0}$ is the growth timescale of the barotropic vorticity under resonance:
\begin{equation}  \label{eq:tau_Z0}
    \tau_{Z0}
    \equiv 2 \Omega \tau^2_*
    \left( \Omega \tau_d + \frac{1}{\Omega \tau_d} \right).
\end{equation}
Equation (\ref{eq:Z0_relaxation_equation}) shows a positive feedback on $\widetilde{Z_0}$: a higher $Z_0$ enhances the inertial stability, making the diabatic heating more efficiently produce $Z_1$ and therefore $Z_0$. The solution to Eq. (\ref{eq:Z0_relaxation_equation}) is:
\begin{equation}  \label{eq:Z0_relaxation_solution}
\begin{split}
    \widetilde{Z_0}
    = f \left[ \exp\left(\frac{t}{\tau_{Z0}}\right) - 1 \right]
    \approx
    \begin{cases} 
    f \frac{t}{\tau_{Z0}}, \quad t \ll \tau_{Z0}, \\
    f \exp \left( \frac{t}{\tau_{Z0}} \right), \quad t \gg \tau_{Z0}.
\end{cases}
\end{split}    
\end{equation}
The $\widetilde{Z_0}$ grows linearly with time at $t \ll \tau_{Z0}$. At $t \gg \tau_{Z0}$, $\widetilde{Z_0}$ grows exponentially due to the enhancement of the inertial stability by $Z_0$. Figures \ref{fig:lag_w0}a and \ref{fig:lag_entrain}a show that the analytical solution [Eq. (\ref{eq:Z0_relaxation_solution})] agrees well with the barotropic component of the simulated vorticity at the domain center. 





\begin{figure}[h]
\centerline{\centering
\includegraphics[width=1\linewidth]{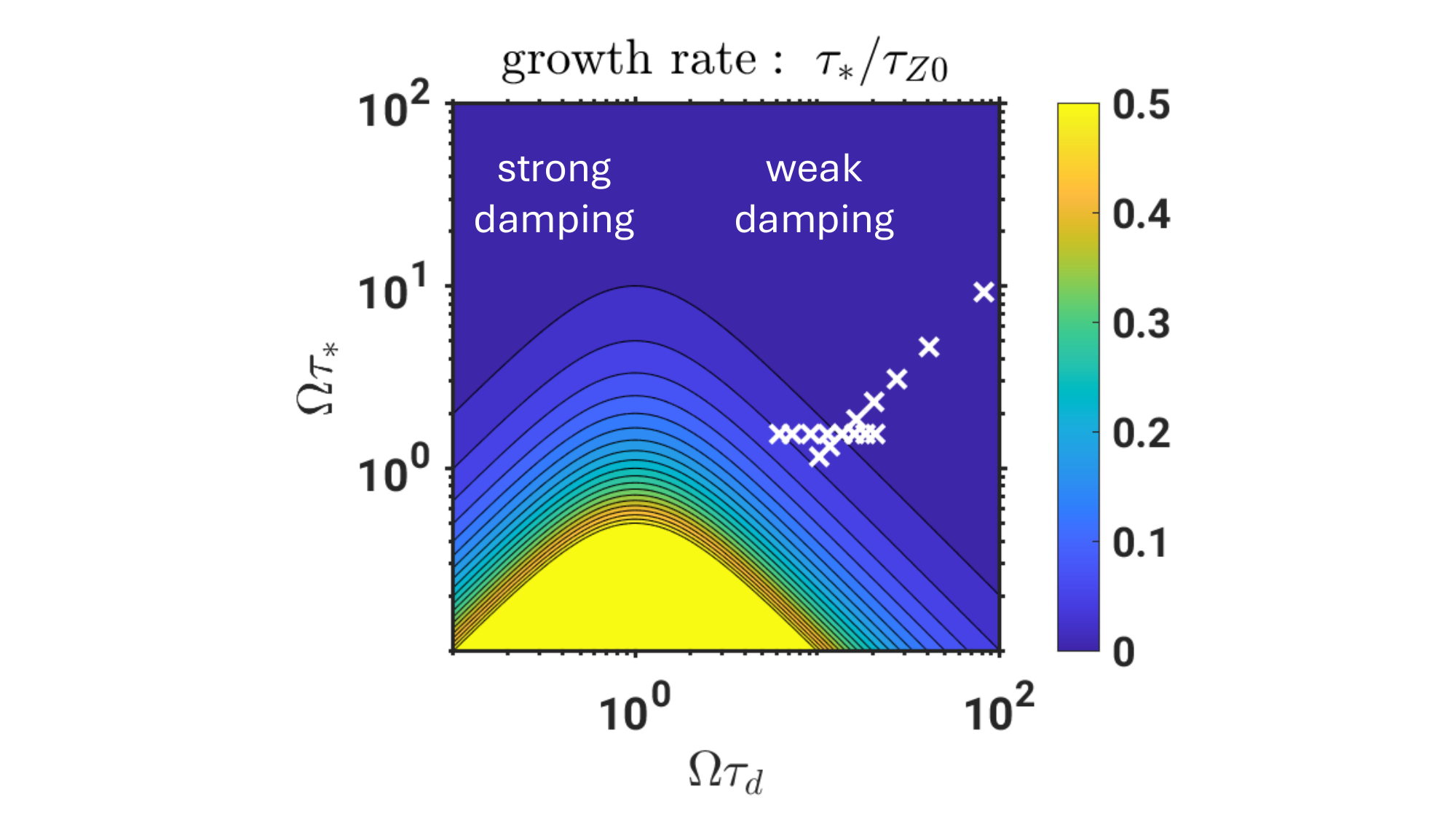}}
  \caption{The growth rate of the barotropic vorticity mode $1/\tau_{Z_0}$ predicted by Eq. (\ref{eq:Z0_relaxation_solution}). It is nondimensionalized by multiplying $\tau_*$. The contour interval is 0.025. The white crosses mark the second and third groups of experiments where CMT is active. Logarithmic coordinates are used. }\label{fig:growth_rate_tao_Z0}
\end{figure}

The growth timescale of barotropic vorticity, $\tau_{Z0}$, is controlled by $\Omega \tau_*$ and $\Omega \tau_d$. The parameter space is shown in Fig. \ref{fig:growth_rate_tao_Z0}:
\begin{itemize}
    \item The growth is accelerated by a lower $\Omega \tau_*$. A greater convective mass flux amplitude $M_*$ reduces the overturning timescale $\tau_*$. It produces a greater amplitude of vorticity ($Z_1$) and a stronger nonlinearity, accelerating the production of barotropic vorticity. 
    \item The growth rate is non-monotonic to $\Omega \tau_d$. For $\Omega \tau_d \ll 1$, the system is in the strong-damping regime where a stronger Rayleigh damping overly dissipates $Z_1$, decelerating the production of barotropic vorticity. For $\Omega \tau_d \gg 1$, the system is in the weak-damping regime where a stronger Rayleigh damping induces a more significant phase shift, accelerating the production of barotropic vorticity. 
\end{itemize}
All of our experiments, which are relevant to the real atmospheric state, lie in the weak-damping regime. We remind readers that this corresponds to the strong-entrainment regime, where a stronger entrainment makes the in-cloud momentum too close to the environment and leads to weaker damping. 

The above analysis focuses on the dynamics at the domain center. Appendix D studies the radial structure of the vortex and points out that the width of the barotropic cyclone is roughly half of the convective region radius $L_M$.

\section{Conclusion}
\label{sec:summary}

Convective activity is crucial for TC genesis, but the relative role of steady versus fluctuating components of convective activity remains unclear. This paper investigates the circulation response to a periodic and local convective activity in an idealized model framework. The periodicity represents the fluctuating components of convective activity in TC genesis, such as the diurnal cycle and inertial-gravity waves. The local convective region represents a mesoscale convective system, which is anticipated to develop into a TC precursor vortex. The primary difference from previous studies is the incorporation of convective momentum transfer (CMT), a frictional, irreversible process that we demonstrate is crucial for the formation of a barotropic cyclonic core. 

We build an axisymmetric numerical model in the Boussinesq approximation. To avoid erroneous representation of convective activity as a convective ring and to retain analytical tractability, we employ a bulk-plume convective parameterization scheme. The convective mass flux is set by a sinusoidal function in time, representing the alternation between the updraft and downdraft phases. Both the dynamical and eddy entrainment/detrainment processes are included. The convective mass flux induces periodic diabatic heating and CMT. When CMT is turned off, no significant barotropic vorticity anomaly is seen in the simulations. When CMT is turned on, a barotropic vorticity anomaly forms, with a cyclonic core wrapped by an anticyclonic shell. 

To explain the formation of barotropic vorticity, a variable-separation analytical strategy is adopted, which separately studies the vertical structure at the vortex core and the radial profile. For the vortex core, we develop a single-column model with a linearized CMT process. The weak temperature gradient approximation is then used to determine the environmental (resolved) vertical velocity in conjunction with the convective mass flux. Truncation to the two leading vertical Fourier modes yields a reduced model, which predicts that a barotropic vortex core first grows linearly with time and then transitions to exponential growth. The basic growth mechanism is the modulation of the phase relation between vertical velocity and the first-baroclinic mode vorticity ($Z_1$). Without CMT, the phases of the two variables are in quadrature. With CMT, their phase lag decreases, resulting in a net generation of barotropic vorticity in each cycle. The transition to exponential growth is due to the amplification of $Z_1$ production rate by the increased inertial stability at the vortex core. The radial structure model shows that the size of the barotropic vorticity region is roughly the same as the convective region, with the inner half of the radius being taken by the cyclonic core and the outer half being taken by the anticyclonic shell. 


The theory predicts that the strength of the cyclonic core depends non-monotonically on the Rayleigh damping timescale of CMT. This is because the damping has dual roles. On the one hand, it increases the phase lag between vertical velocity and $Z_1$, thereby enhancing barotropic vorticity production. On the other hand, the damping suppresses the amplitude of $Z_1$, disfavoring the production of barotropic vorticity. In the weak-damping regime, the phase modulation effect dominates; in the strong-damping regime, the amplitude suppression effect dominates. The real atmospheric state is estimated to be in the weak-damping regime. This turns out to be in the strong entrainment regime, where stronger entrainment makes the in-cloud wind too close to the environmental wind, inducing less CMT.  

This research has several limitations that need to be emphasized. First, the model and theory neglect surface friction. The CMT is an internal transport mechanism in the atmosphere that only acts on baroclinic modes, but surface friction influences all modes \cite[e.g.,][]{neelin2000quasi,wu2000rayleigh}. Thus, surface friction may reduce the significance of barotropic vorticity growth. Second, regarding the convective parameterization, a simple bulk plume scheme is employed. The plume adjustment time is assumed to be infinitely short, neglecting the convective memory effect \citep{pan1998cumulus,davies2009memory,mapes2011parameterizing,colin2021atmospheric}. The diabatic heating due to stratiform precipitation is also neglected. The horizontal pressure gradient force is essentially ignored. Thus, the current convective parameterization is still very crude for a mesoscale convective system. Third, the time evolution of convective mass flux is prescribed as a sinusoidal function in time. Still, the fluctuation of convective activity in the real atmosphere involves significant irregularity \cite[e.g.,][]{ooyama1982conceptual,van2008predictability,biagioli2023dimensionless,fu2024stochastic,fu2025quasi}. Thus, noise, which contains a spectrum of timescales, can be added to the time evolution of the convective mass flux. One can examine whether the barotropic vorticity still builds up. 

In conclusion, this paper employs an axisymmetric model to demonstrate that the two components of periodic convective activity, diabatic heating and CMT, cooperate to produce a barotropic vorticity structure. To accelerate TC genesis, the CMT does \textit{not} need to be upgradient: the downgradient momentum transfer can accelerate the vortex growth by modulating the phase relation between $w$ and $\zeta$. The mechanism is \textit{not} a resonance: a fluctuation with any period applies. The theory implies that a persistent convective activity may \textit{not} be the sole route for vortex development. For future work, a 3D cloud-resolving model is needed to confirm that the fluctuation of convective activity, either due to the diurnal cycle or inertial-gravity waves, can indeed accelerate TC genesis.







\acknowledgments
Preliminary results regarding the asymptotic analysis of the reduced model have been reported in Hao Fu's Ph.D. thesis at Stanford University. The axisymmetric model and the simplified convective parameterization scheme were developed during Hao Fu's tenure at the University of Chicago as a T. C. Chambelrin Postdoctoral fellow and at Nanjing University as a faculty member. The author benefits from discussions with Christopher A. Davis at NCAR, Morgan E O'Neill at the University of Toronto, and Bolei Yang, Han Chen, and Yue Shen at Nanjing University. Hao Fu acknowledges the NCAR CISL lab for providing computational resources, with kind support from Da Yang at Stanford University.


%
%
\datastatement
The movie version of Fig. 2 for all experiments, the axisymmetric model, the postprocessing codes, and a mathematical derivation note can be downloaded from: https://box.nju.edu.cn/d/653d1189c69c40f8be58/.





\appendix[A]    \label{app:discretization}
\appendixtitle{Discretization of the axisymmetric model}

\begin{figure}[h]
\centerline{\centering
\includegraphics[width=0.6\linewidth]{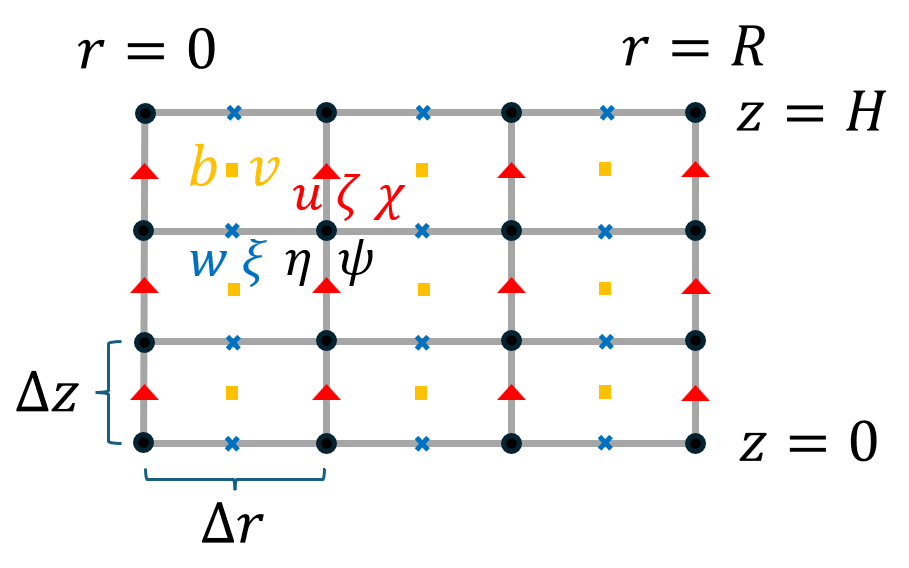}}
  \caption{A schematic diagram for the staggered grid of the axisymmetric model. }\label{fig:grid}
\end{figure}

The model is discretized on a staggered grid with the finite-difference method. The domain is split into $N_r\times N_z$ cells, with $N_r$ cells in the radial direction and $N_z$ cells in the vertical direction. The radial grid spacing is $\Delta r$, and the vertical grid spacing is $\Delta z$. See Fig. \ref{fig:grid} for an illustration of the grid. 
\begin{itemize}
    \item The $\eta$ and $\psi$ are defined at the grid corners (black circles).
    \item The $b$, $v$, and $v_c$ are defined at the grid centers (orange squares).
    \item The $u$, $u_c$, and $\zeta$ are defined on the grid columns (red triangles).
    \item The $w$, $\xi$, $\epsilon$, and $\delta$ are defined on the grid rows (blue crosses). 
\end{itemize}
The flux terms in the $\eta$, $\zeta$, and $b$ equations are calculated as the difference of flux values. The flux values use quantities linearly interpolated from neighboring grid points. The time-stepping of Eqs. (\ref{eq:eta})-(\ref{eq:b}) uses the second-order Adams-Bashforth scheme \citep{durran2010numerical}. 

The streamfunction equations [(\ref{eq:chi}) and (\ref{eq:psi})] are solved with a combination of finite-difference and Fourier transform methods. For the $\chi$ equation (\ref{eq:chi}), which is 1D, the finite difference in the radial direction is used. For the $\psi$ equation (\ref{eq:psi}), which is 2D, we employ the classic method of \citet{hockney1965fast} to solve the finite-difference problem with a Fourier transform. The domain is first extended to $0 \le z \le 2H$ with antisymmetry about $z=H$. Then, the Fourier transform is performed. The governing equations of the Fourier coefficients are discretized with the finite-difference method in the radial direction, leading to a tridiagonal matrix of the Fourier coefficients of $\psi$. Finally, we perform an inverse Fourier transform to obtain $\psi$.

\appendix[B]
\appendixtitle{Derivation of the bulk-plume model}

This appendix follows \citet{romps2012equivalence} to review the derivation of the bulk-plume model. The convective fractional area is referred to as $\sigma_c$, and the in-cloud vertical velocity is referred to as $w_c$. The convective mass flux is defined as:
\begin{equation}  \label{eq:M_sigma_c}
    M \equiv \rho \sigma_c w_c.
\end{equation}
The in-cloud wind speed is ($u_c$, $v_c$), and the environmental wind speed is ($u$, $v$). The entrainment and detrainment rate, $e$, and $d$ (unit: kg m$^{-3}$ s$^{-1}$), are related to the fractional entrainment and detrainment rates via:
\begin{equation}  \label{eq:e_d_epson_delta}
    \epsilon \equiv \frac{e}{M}, \quad \delta \equiv \frac{d}{M}.
\end{equation}
The mass continuity for the plume and environmental parts obeys:
\begin{equation} \label{eq:continuity_c}
    \underbrace{ \frac{\partial}{\partial t} \left( \sigma_c \rho \right) }_{\approx 0} + \frac{\partial}{\partial z} \underbrace{ \left(  \sigma_c \rho w_c \right) }_{M} = e - d,
\end{equation}
\begin{equation} \label{eq:continuity_e}
    \frac{\partial}{\partial t} \left( \sigma_e \rho \right) + \frac{\partial}{\partial z} \left( \sigma_e \rho w \right) = d - e.
\end{equation}
where $\sigma_e \equiv 1 - \sigma_c$ is the environmental fractional area. The tangential momentum conservation law for the plume and environmental parts obeys:
\begin{equation} \label{eq:momentum_c}
    \underbrace{ \frac{\partial}{\partial t} \left( \sigma_c \rho v_c \right) }_{\approx 0} + \frac{\partial}{\partial z} \left( \sigma_c \rho w_c v_c \right) = e v - d v_c + F,
\end{equation}
\begin{equation} \label{eq:momentum_e}
    \frac{\partial}{\partial t} \underbrace{ \left( \sigma_e \rho v \right) }_{\approx \rho v} + \frac{\partial}{\partial z} \left( \sigma_c \rho w_c v \right) = d v_c - e v - F,
\end{equation}
where $F$ is the horizontal pressure gradient force imposed on the in-cloud $v$ momentum. According to Newton's third law, a force with a minus sign appears in the environmental momentum equation. The $u$ momentum equation has the same structure, so it is not explicitly shown.  

Two assumptions will be used:
\begin{enumerate}
    \item $\sigma_c \ll 1$, which indicates $\sigma_e \approx 1$ in Eqs. (\ref{eq:continuity_e}) and (\ref{eq:momentum_e}).
    \item Fast adjustment of the in-cloud mass and momentum, which removes the tendency term in Eqs. (\ref{eq:continuity_c}) and (\ref{eq:momentum_c}). To understand this, we note that the right-hand side terms of Eqs. (\ref{eq:continuity_c}) and (\ref{eq:continuity_e}) have comparable magnitudes. However, $\sigma_c \ll 1$ indicates the tendency term in Eqs. (\ref{eq:continuity_c}) and (\ref{eq:momentum_c}) should be much smaller than other terms. 
\end{enumerate}


With these assumptions, Eq. (\ref{eq:continuity_c}) reduces to Eq. (\ref{eq:epson_sum}). Substituting Eq. (\ref{eq:continuity_c}) into Eq. (\ref{eq:momentum_c}), we obtain the in-cloud momentum equations [(\ref{eq:uc_updraft}) and (\ref{eq:uc_downdraft})]. Substituting Eqs. (\ref{eq:M_sigma_c})-(\ref{eq:momentum_c}) into Eq. (\ref{eq:momentum_e}), we obtain the environmental momentum equation (\ref{eq:dudt}).

\appendix[C]    \label{app:damping}
\appendixtitle{The Rayleigh damping timescale}

In this appendix, we follow \citet{romps2014rayleigh} to derive the Rayleigh damping timescale associated with the convective momentum transfer, using the bulk-plume equation. The updraft phase, which is studied by \citet{romps2014rayleigh}, is used as an example. The Rayleigh damping for the downdraft phase has not been studied before. At the end of this appendix, we apply a symmetry assumption to extend the solution from the updraft to the downdraft phase, which yields the same $\tau_d$. 

Because the CMT parameterization is vertically 1D, we derive the Rayleigh damping on $v$ as an example. The same law holds for $u$. The following assumptions are used to linearize the problem:
\begin{enumerate}
    \item The convective mass flux $M$ is vertically uniform.
    \item The dynamical entrainment and detrainment are neglected ($\epsilon_{dyn} = \delta_{dyn} = 0$), and the eddy entrainment and detrainment are equal and constant ($\epsilon_{edd}=\delta_{edd} = \mathrm{const}$). 
    \item The boundary condition is assumed cyclic in the $z$ direction.
\end{enumerate}
With these simplifications, the mass flux equation reduces to a linear one:
\begin{equation}  \label{eq:vc_simplified}
    \frac{\partial v_c}{\partial z} = \epsilon (v-v_c),
\end{equation}
\begin{equation}  \label{eq:v_simplified}
    \rho \frac{\partial v}{\partial t} = M \frac{\partial (v - v_c) }{\partial z}.
\end{equation}
To focus on CMT, other right-hand side terms of the $v$ equation are temporarily neglected. Because Eqs. (\ref{eq:vc_simplified}) and (\ref{eq:v_simplified}) are linear equations with constant coefficients, we substitute in a normal mode solution of $v_c$ and $v$:
\begin{equation}  \label{eq:normal_mode}
\begin{split}
   v_c &= \hat{v}_c \exp \left[ i \frac{\pi}{H} z - \left( \frac{1}{\tau_d} + i \omega \right) t \right], \\
   v &= \hat{v} \exp \left[  i \frac{\pi}{H} z- \left( \frac{1}{\tau_d} + i \omega \right) t \right],
\end{split}   
\end{equation}
where $1/\tau_d$ is the real part of the growth rate, and $-i\omega$ is the imaginary part of the growth rate. The imaginary part indicates a traveling disturbance in the vertical direction. Here, $\tau_d$ is the Rayleigh damping timescale, $\omega$ is the phase shift rate of $v_c$ and $v$ (not to be confused with the angular frequency of the periodic convection $\Omega$), and $\hat{v}_c$ and $\hat{v}$ are the complex amplitudes. Substituting Eq. (\ref{eq:normal_mode}) into Eqs. (\ref{eq:v_simplified}) and (\ref{eq:vc_simplified}), we obtain a linear algebraic equation set:
\begin{equation}
   \begin{bmatrix}
   i \frac{\pi}{H} + \epsilon  & -\epsilon \\ 
   i M \frac{\pi}{H} & - \rho \left( \frac{1}{\tau_d} + i \omega \right) - i M \frac{\pi}{H} \\
   \end{bmatrix} 
   \begin{bmatrix}
   \hat{v}_c \\ 
   \hat{v} \\
   \end{bmatrix} 
   = 0.
\end{equation}
The existence of a solution requires the determinant of the matrix to be zero for both the real and imaginary parts. This yields:
\begin{equation}  \label{eq:tao_d_omega}
\begin{split}
    \tau_d &= \frac{\rho}{M}\frac{\epsilon^2+\pi^2/H^2}{\epsilon \pi^2/H^2}, \\
    \omega &= - \frac{1}{\tau_d} \frac{\pi}{\epsilon H}.
\end{split}    
\end{equation}
The $\tau_d$ expression here is used as Eq. (\ref{eq:tau_Rayleigh}). In the updraft phase, $M$ is positive, yielding a positive $\tau_d$ and a negative $\omega$. This indicates that the free-tropospheric wind profile is not only damped but also descends, as noted by \citet{romps2014rayleigh}. Given the fact that the vertical wavenumber is $\pi/H$, the descending speed is:
\begin{equation}
    c_z = \frac{\omega}{\pi/H} = - \frac{M}{\rho} \underbrace{ \left( \epsilon^2 \frac{\pi^2}{H^2} + 1 \right)^{-1} }_{<1},
\end{equation}
which is slower than the compensating descent speed $M/\rho$. This indicates that the descent is primarily due to the vertical advection of the wind profile by the descending flow; however, the lateral exchange between the plume and the environment slows it down.

In the downdraft phase, $M$ is negative, but there is an additional minus sign at the right-hand side of Eq. (\ref{eq:vc_simplified}). The $\tau_d$ remains positive, but $\omega$ changes to positive, indicating an upward propagating perturbation.

\appendix[D]    \label{app:radial_structure}
\appendixtitle{The radial structure of the barotropic vorticity component}

\section{Qualitative analysis}

The main body of the text primarily examines the formation of a barotropic cyclone at the domain center. In this appendix, we quantitatively study the radial distribution of barotropic vorticity in an axisymmetric vortex, which determines the width of the barotropic cyclone. The analysis shows that an anticyclonic shell wraps the barotropic cyclone, and the transitional radius is roughly half the width of the convective region $L_M$. 

The advection form of the vertical vorticity equation (\ref{eq:zeta}), neglecting the eddy viscosity and the sponge layer damping term, reads:
\begin{equation}   \label{eq:vort_axis}
   \frac{\partial \zeta}{\partial t} + {u}  \frac{\partial \zeta}{\partial r} + w \frac{\partial \zeta}{\partial z} 
   \approx - \frac{\partial v}{\partial z} \frac{\partial w}{\partial r} + \left( f + \zeta \right) \frac{\partial w}{\partial z} + C_\zeta. 
\end{equation}
Compared to the vorticity equation depicting the domain center [Eq. (\ref{eq:zeta_center})], Eq. (\ref{eq:vort_axis}) has an additional radial advection term ${u}\partial \zeta/\partial r$ and a tilting term $-(\partial v/\partial z)(\partial w/\partial r)$. The radial advection term makes a cyclone more compact at the convergent level and an anticyclone less compact at the divergent level. At the updraft phase, the updraft tilts the radially outward-pointing horizontal vortex tube downward; at the downdraft phase, the downdraft tilts the radially inward-pointing horizontal vortex tube downward. Therefore, both the radial advection and tilting terms generate an anticyclonic shell that wraps the anticyclonic core. Such a core-shell barotropic vorticity is possibly a universal feature of convection in rotating fluids, as it has been reported not only in simulated TC precursor vortices \citep{tory2006prediction_part2} but also in rotating Rayleigh-Bénard convection \citep{fu2024RRB}.  

The core-shell barotropic vorticity structure has complicated dynamical implications, as reported in the literature of vortex dynamics. It prevents the cyclonic core from inducing an anomalous far-field cyclonic flow, which suppresses vortex interaction \cite[e.g.,][]{li2020shield}. Nevertheless, the straining flow induced by neighboring vortices may peel off the anticyclonic shell \cite[e.g.,][]{carton1992merger}. This appendix focuses on the quantitative depiction of the core-shell structure, which has not been made before.

\section{Quantitative analysis}

To quantify the radial distribution of barotropic vorticity, we introduce an asymptotic argument, which decomposes $w$ and $\zeta$ by vertical Fourier modes:
\begin{equation}
    w = w_1 + w_0 + ..., \quad \zeta = \zeta_1 + \zeta_0 + ...,
\end{equation}
where $w_1$ has a $\sin(\pi z/H)$ vertical structure, $w_0$ is zero; $\zeta_1$ has a $\cos(\pi z/H)$ vertical structure, and $\zeta_0 \equiv \overline{\zeta}$ is vertically uniform. Assuming the $n=0$ mode quantities have a much smaller magnitude than the $n=1$ mode ones, we adopt an asymptotic approach: solving for the $\zeta_1$ first and then $\zeta_0$.

Given the WTG approximation [Eq. (\ref{eq:w_WTG})], $w_1$ is largely determined by $Q$ and therefore the prescribed convective activity. Thus, $w_1$ has a variable-separation form:
\begin{equation}  \label{eq:variable_separation}
    w_1 = \mathcal{R}(r) \Phi(z) T_w (t).
\end{equation}
Here, $\mathcal{R}(r)$ is the radial structure (not to be confused with the potential radius), which obeys a Gaussian function:
\begin{equation}  \label{eq:R_Gaussian}
    \mathcal{R}(r) = \exp \left( - \frac{r^2}{L_M^2} \right),
\end{equation}
$\Phi(z)$ is the vertical structure:
\begin{equation}
 \Phi(z) = \sin \left( \frac{\pi z}{H} \right),   
\end{equation}
and $T_w(t)$ is the time evolution part that carries the unit (m s$^{-1}$), which needs to be solved from the $n=1$ mode vorticity equation:
\begin{equation}  \label{eq:vort_axis_0}
   \frac{\partial \zeta_1}{\partial t} + \frac{\zeta_1}{\tau_d} = f \frac{\partial w_1}{\partial z} = f \mathcal{R} \frac{d\Phi}{dz} T_w. 
\end{equation}
The $\tau_d$, which depends on $M$, is a function of radius. To make the problem analytically tractable, we make a strong assumption here: \textit{$\tau_d$ is approximately not a function of radius}. At the end of this appendix, this assumption is speculated to severely limit the accuracy of the theory. 

Letting $\zeta_1$ be zero at $t=0$, the solution of Eq. (\ref{eq:vort_axis_0}) is:
\begin{equation}
    \zeta_1 = f \mathcal{R} \frac{d\Phi}{dz} \underbrace{ \int_0^t T_w(t') \exp \left( -\frac{t-t'}{\tau_d} \right) dt' }_{T_\zeta(t)},
\end{equation}
where $T_\zeta(t)$ denotes the time-dependent part of $\zeta_1$, and $t'$ is a dummy integration variable. The $\zeta_1$ can be used to calculate the $n=1$ mode tangential velocity $v_1$ and radial velocity $u_1$:
\begin{equation}   \label{eq:u_theta}
    v_1 = \frac{1}{r} \int_0^r \zeta_1 rdr = f T_\zeta \frac{d\Phi}{dz} \frac{1}{r} \int_0^r \mathcal{R} rdr,
\end{equation}
\begin{equation}   \label{eq:{u}}
    u_1 = -\frac{1}{r} \int_0^r \frac{\partial w_1}{\partial z} rdr = - T_w \frac{d\Phi}{dz} \frac{1}{r} \int_0^r \mathcal{R} rdr.
\end{equation}
We remark that the assumption of uniform $\tau_d$ in the radial direction is crucial in making $\zeta_1$, $v_1$, and $u_1$ exhibit a variable-separation form, which simplifies the subsequent analysis.

With the above preparations, we derive the $n=0$ mode vorticity equation, using the vertical average of Eq. (\ref{eq:vort_axis}):
\begin{equation}  \label{eq:zeta_1_appendix_p_radial}
\begin{split}
    \frac{d {\zeta_0}}{dt} + \frac{{\zeta_0}}{\tau_d} 
    &= \overbrace{ -\overline{\Phi \frac{d^2\Phi}{dz^2}} f T_w T_\zeta \mathcal{R}^2 }^{\text{from }-w_1 \frac{\partial \zeta_1}{\partial z}} 
 \,\,\,
    + \,\,\,\overbrace{ \overline{ \left( \frac{d\Phi}{dz} \right)^2 } f T_w T_\zeta \mathcal{R}^2 }^{\text{from }\zeta_1 \frac{\partial w_1}{\partial z}} \\
    &\underbrace{ - \overline{\Phi \frac{d^2\Phi}{dz^2}} f T_w T_\zeta \frac{d\mathcal{R}}{dr} \frac{1}{r}\int_0^r \mathcal{R} rdr }_{\text{from }- \frac{\partial v_1}{\partial z}\frac{\partial w_1}{\partial r} }
    \,\,\,+\,\,\, \underbrace{ \overline{\left( \frac{d\Phi}{dz} \right)^2 } f T_w T_\zeta \frac{d\mathcal{R}}{dr} \frac{1}{r}\int_0^r \mathcal{R} rdr }_{\text{from }-{u}_1\frac{\partial \zeta_1}{\partial r}  } \\
    &= \underbrace{ 2 \overline{\left( \frac{d\Phi}{dz} \right)^2 } }_{=1} f T_w T_\zeta \underbrace{ \left(  \mathcal{R}^2 
 + \frac{d\mathcal{R}}{dr}\frac{1}{r}\int_0^r \mathcal{R}rdr\right) }_{ \mathcal{R}_{\zeta_0}}, 
\end{split}    
\end{equation}
where the overbar denotes calculating the vertical average. The vertical average of $f\partial w_1/\partial z$ vanishes due to the non-penetration boundary condition of $w_1$. Vertical integration by parts is employed to merge the right-hand side terms. Here, $\mathcal{R}_{\zeta_0}$ represents the radial structure of the right-hand side terms of Eq. (\ref{eq:zeta_1_appendix_p_radial}), which determines the radial structure of the barotropic vorticity $\zeta_0$.

\begin{figure}[h]
\centerline{\centering
\includegraphics[width=1\linewidth]{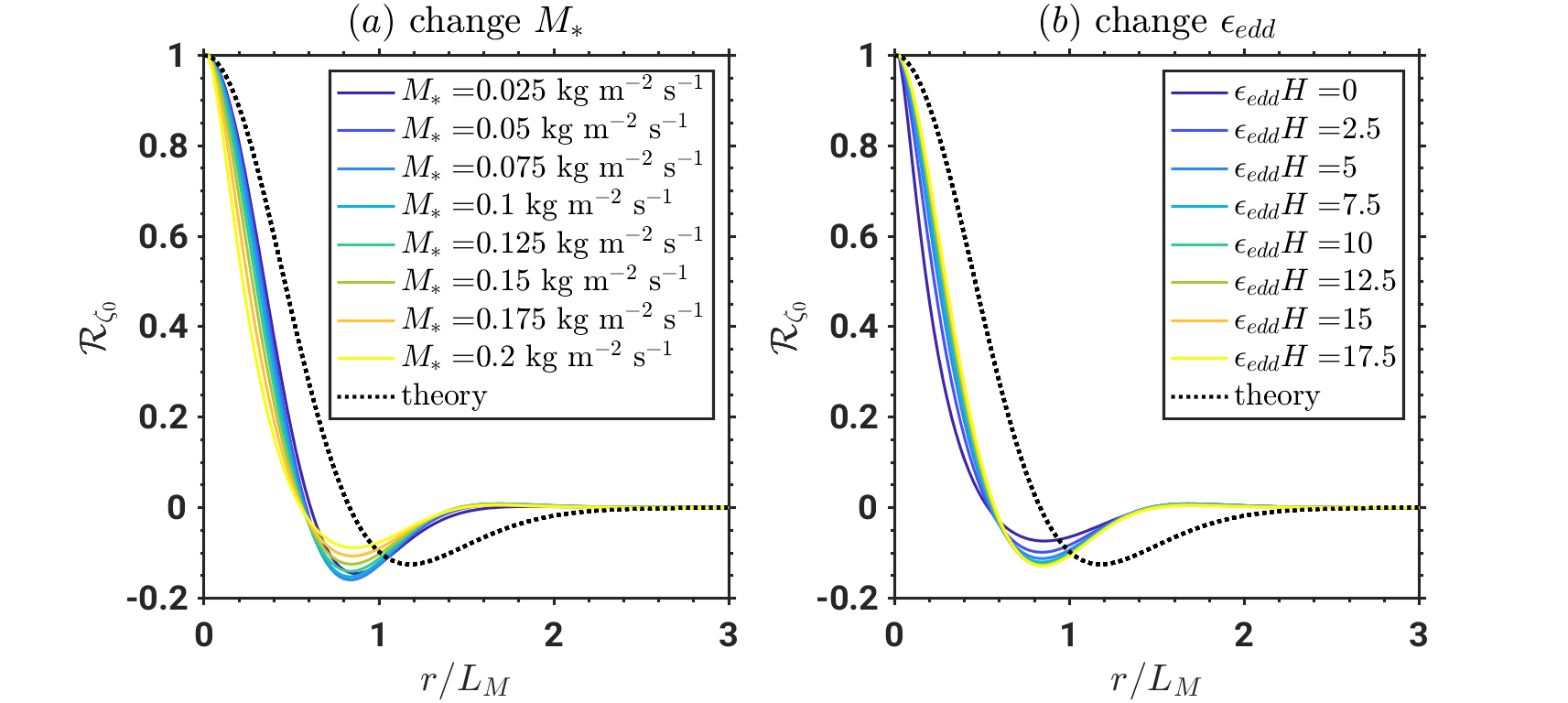}}
  \caption{(a) The colored lines show the radial structure of the barotropic vorticity, $\overline{\zeta}/\overline{\zeta}|_{r=0}$, for the eight CMT\_on experiments. Colors from dark to bright show experiments with increasing $M_*$. The dotted black line shows the theoretical prediction [Eq. (\ref{eq:Iz1_radial_ratio})]. (b) The same as (a), but for the eight experiments in the EDD group that change the eddy fractional entrainment rate. Colors from dark to bright show the results of experiments with increasing $\epsilon_{edd}$.  }\label{fig:radial_Z0}
\end{figure}   

Substituting Eq. (\ref{eq:R_Gaussian}) into Eq. (\ref{eq:zeta_1_appendix_p_radial}), we obtain the expression of $\mathcal{R}_{\zeta_0}$:
\begin{equation}
\begin{split}
\label{eq:Iz1_radial_ratio}
   \mathcal{R}_{\zeta_0} \equiv \frac{{\zeta_0}}{{\zeta_0}|_{r=0}} 
    &= \overbrace{ - \exp \left(-\frac{r^2}{L_M^2} \right) + \exp\left( -\frac{2r^2}{L_M^2} \right) }^{\frac{d\mathcal{R}}{dr}\frac{1}{r}\int_0^r \mathcal{R}rdr} 
    + \overbrace{ \exp \left( -\frac{2r^2}{L_M^2} \right) }^{\mathcal{R}^2} \\
    &= - \exp \left( -\frac{r^2}{L_M^2} \right) + 2 \exp \left( -\frac{2r^2}{L_M^2} \right),
\end{split}    
\end{equation}
whose zero point corresponds to the transition radius of cyclone to anticyclone $r_{\mathrm{CC \rightarrow AC}}$:
\begin{equation}
   \label{eq:r_CC_AC}
    r_{\mathrm{CC \rightarrow AC}} = (\ln 2)^{1/2} L_M \approx 0.83 L_M.
\end{equation}
The transition radius is approximately the \textit{halfwidth} of a Gaussian-shaped updraft.

Figure \ref{fig:radial_Z0} plots the radial structure of barotropic vorticity $\zeta_0$ diagnosed from the axisymmetric simulations, as well as the theoretical prediction [Eq. (\ref{eq:Iz1_radial_ratio})]. Both the simulation and the theory exhibit a core-shell vorticity structure, as evidenced by an overshoot of $\zeta_0$ in the radial direction. However, the theory overestimates the width of the overshooting structure by roughly 30\%. We speculate that this overestimation is due to the assumption of uniform $\tau_d$ in the radial direction. As we move outward from the domain center, $M$ decreases, $\tau_d$ increases (Fig. \ref{fig:Nz}b), indicating that the Rayleigh damping gets weaker. Because the Rayleigh damping is crucial for producing barotropic vorticity under periodic convective activity, we speculate that a weaker Rayleigh damping in the outer region makes the core-shell structure more concentrated in the domain's central region.


\bibliographystyle{ametsocV6}
\bibliography{references}

\end{document}